\documentclass[reprint,showkeys,nofootinbib,amsmath,amssymb,aps,10pt,superscriptaddress,onecolumn
]{revtex4-2}

\usepackage[T1]{fontenc}
\usepackage[utf8]{inputenc}
\usepackage{url}

\usepackage{graphicx}
\usepackage{xcolor}
\usepackage{multirow}
\definecolor{darkblue}{RGB}{0,0,196}
\usepackage[colorlinks=true,linktocpage=true,linkcolor=darkblue,citecolor=red,urlcolor=darkblue]{hyperref}
\usepackage{caption}
\usepackage[margin=2.5cm]{geometry}

\usepackage{float}
\usepackage{caption}
\usepackage{subcaption}

\usepackage[strings]{underscore} 

\usepackage[normalem]{ulem}

\begin{document}

\title{Multifractal Signatures of Ageing and Dementia Development: A Multifractal Space-Filling Curve Analysis}

\author{Marta Lotka}
\email[(corresponding author) ]{marta.lotka@doctoral.uj.edu.pl}
\affiliation{Faculty of Physics, Astronomy and Applied Computer Science, Jagiellonian University, 30-348 Kraków, Poland}
\affiliation{Doctoral School of Exact and Natural Sciences, Jagiellonian University, 30-348 Kraków, Poland}
\author{Jacek Grela}
\email[(corresponding author) ]{jacekgrela@gmail.com}
\affiliation{Institute of Theoretical Physics, Jagiellonian University, 30-348 Kraków, Poland}
\affiliation{Mark Kac Center for Complex Systems Research, Jagiellonian University, 30-348 Kraków, Poland}
\author{Zbigniew Drogosz}
\email[(corresponding author) ]{zbigniew.drogosz@alumni.uj.edu.pl}
\affiliation{Institute of Theoretical Physics, Jagiellonian University, 30-348 Kraków, Poland}
\author{Jeremi K. Ochab}
\email[(corresponding author) ]{jeremi.ochab@uj.edu.pl}
\affiliation{Institute of Theoretical Physics, Jagiellonian University, 30-348 Kraków, Poland}
\affiliation{Mark Kac Center for Complex Systems Research, Jagiellonian University, 30-348 Kraków, Poland}
\author{Paweł Oświęcimka}
\email[(corresponding author) ]{pawel.oswiecimka@ifj.edu.pl}
\affiliation{Mark Kac Center for Complex Systems Research, Jagiellonian University, 30-348 Kraków, Poland}
\affiliation{Complex Systems Theory Department, Institute of Nuclear Physics, Polish Academy of Sciences, 31-342 Kraków, Poland}

\begin{abstract}
Multifractality is an effective formalism for quantifying the nonlinear, scale-free properties of complex data. In this study, we propose a novel and efficient methodology, termed Multifractal Space-filling Curve Analysis (MFSCA), for quantifying the correlation structure of multidimensional data. Within this framework, the original multidimensional data - while preserving both local and long-range organisational properties - are projected onto a one-dimensional representation using a fractal space-filling curve. The resulting one-dimensional signal is then analysed using multifractal algorithms. We demonstrate the utility of the method using both artificially generated multifractal structures and real data. In particular, we apply MFSCA to analyse magnetic resonance imaging (MRI) data from Alzheimer patients at different stages of dementia. Based on the results, we estimate the multifractal profiles of the brain for healthy subjects of different ages as well as for dementia patients. The analysis reveals that the spatial organization of brain structures, as measured by the degree of multifractality, progressively weakens with age and the development of dementia. A transition from multifractality to monofractality is observed both in control groups, when comparing the Young Control and Elderly Control groups, and among dementia subjects of similar age but at different stages of the disease, namely early dementia and mild cognitive impairment. Thus, from the perspective of multiscaling properties, the heterogeneous characteristics of spatial brain organization deteriorate under worsening conditions, leading to a homogeneous and weakly correlated structure. These findings not only effectively capture key aspects of brain organisation, but also demonstrate that the multifractality of MRI data can serve as a marker of structural brain changes.

\end{abstract}


\maketitle

\section{Introduction}

In 2019, an estimated 57.4 million people were living with dementia globally. This number is expected to increase at an accelerating rate: to 83 million in 2030, 116 million in 2040, and about 152 million by 2050~\cite{alzheimer2019world,nichols2022estimation}. Alzheimer’s disease (AD) is the most common form of dementia, accounting for 60–80\% of all cases~\cite{alzheimer2021world}. The Alzheimer's Association defines AD based on its biology, beginning with the appearance of neuropathologic changes in asymptomatic individuals. Those changes are detected via (early-changing) Core 1 biomarkers. Those include amyloid positron emission tomography (PET) as well as approved cerebrospinal fluid (CSF) and plasma biomarkers focused on amyloid-beta (A$\beta$) proteinopathy or on phosphorylated and secreted AD tau proteins.
Later-changing Core 2 biomarkers, including tau PET and additional tau-oriented biofluid analyses, provide additional information~\cite{jack2018nia,jack2024revised}.
The longest-established core biomarkers (CSF essays and PET imaging) have significant limitations that preclude them from serving as a means of large-scale population screening. CSF assays are invasive and carry the risk of infection and side effects, whereas PET is costly, and radiation is harmful at high doses~\cite{sperling2011toward}. Novel plasma biomarkers are perceived as more affordable and less invasive,
with potential for more widespread use and more frequent acquisition~\cite{aschenbrenner2022comparison,therriault2023equivalence,noda2024cost,scholl2025cutting,kubota2025plasma,palmqvist2025plasma}.

Progress, however, may come from alternative neuroimaging modalities.
In a recent review,  main shortcomings identified in task-based and resting-state fMRI in the early detection of AD were ``being intensive to collect, clean, preprocess, and analyze data'', and of structural MRI ``rare use of quantification software in clinical practice''~\cite{dang2023neuroimaging}. It follows that there is a need for developing simpler, routine, and more robust MRI and fMRI data analysis methods.
Such advancements could make this imaging modality a more precise and widely used tool for AD prognosis and even lead to its future inclusion in AD diagnostic criteria.

Some of the earlier fMRI studies in AD focused on default-mode network (DMN) differences between AD patients and normal controls (NC), detected using independent component analysis (ICA) \cite{greicius2004default,wu2011altered}. MRI volumetry is another frequently used approach \cite{schuff2009mri,struyfs2020automated,Bachli2020EvaluatingApproach}. Current literature often focuses on machine learning (ML) models \cite{doering2024mri,karim2024identifying}. The recent review \cite{aghdam2025machine} identifies several barriers to clinical adoption of ML models. These barriers include a lack of interpretability, difficulty generalizing across diverse datasets, the scarcity of labeled neuroimaging data (which hinders robust ML training), and insufficiently user-friendly tools that require technical expertise.

The lack of interpretability of ML methods, often regarded as ``black boxes'', motivates the development of more transparent approaches. The application of fractal methods is a promising, theoretically grounded, yet still insufficiently studied pathway. Most studies applying fractal analysis to neurodegenerative disease detection or prediction use a simple measure: the brain's fractal dimension (FD). These methods show promise for early and pre-symptomatic disease detection. However, the limited literature and methodological diversity cause uncertainty regarding their clinical utility~\cite{ziukelis2022fractal}. Even more scarce is the literature on the application of advanced fractal methods and metrics, including Multifractal Detrended Moving Average (MF-DMA) \cite{eke2012pitfalls,rohini2020differentiation} and the Hurst exponent \cite{long2018brainnetome,long2023identifying}. The last two studies combine the approach with support vector machine (SVM) and artificial neural network (ANN) classifiers.

Brain structure changes throughout human lifespan in healthy aging process as well. This needs to be taken into account in prediction of AD, although age prediction from MRI or EEG is a standalone research topic as well.
Here, too, the fractal measure used by most studies is FD, including generalized Rényi dimensions \cite{reishofer2018age}. It has been used for predicting age in adults \cite{madan2016cortical,madan2018predicting,farahibozorg2015age} and in newborns, as the brain undergoes rapid changes in the neonatal period \cite{krohn2026fractal}. Studies correlating fractal features beyond FD with age are much more scarce \cite{dong2018hurst, stanyard2024aperiodic,berthouze2010human}. Some works use them as input for ML techniques \cite{davoudi2025electroencephalography,cauzzo2026detrended}.

Thus, although fractal methods are interesting from the perspective of predicting and monitoring the development of Alzheimer’s disease (AD) or ageing processes, the information provided by these methods is mainly restricted to linear dependencies. This is one of the reasons why multifractality has emerged as a promising tool for the quantitative description of complex data. Its ability to quantify the scale-free properties of heterogeneous data makes this methodology one of the principal tools used in the analysis of data complexity. A serious drawback, however, is its application in the context of multidimensional data. This problem is particularly relevant in the field of neuroscience, where neuroimaging techniques deliver data in more than one dimension, making it difficult to effectively apply multifractal analysis to such datasets. In this work, we address this issue by combining multifractal analysis with space-filling curve techniques, called Multfractal Space-filling Curve Analysis (MFSCA) thereby creating a powerful methodology for the analysis of complex multidimensional data.


The organisation of the paper is as follows. In Section \ref{MFSCA}, the proposed method (MFSCA) is described in detail. Section \ref{methods-description} contains a description of the dataset of Alzheimer’s disease patients analysed in this study, together with the statistical methods used to verify the reliability of the results. Section \ref{Results} presents the results obtained for artificially generated data, namely deterministic two-dimensional multifractal cascades (\ref{sec:cascades}), as well as the results for healthy subjects of different ages and Alzheimer’s disease patients at different stages of dementia (\ref{sec:results_AD}). Section~\ref{sec:conclusions} presents a summary of the results and the final conclusions.

\section{Multi-Fractal Space-filling Curve Analysis (MFSCA)}
\label{MFSCA}

\begin{figure}[ht]
\includegraphics[scale=0.4]{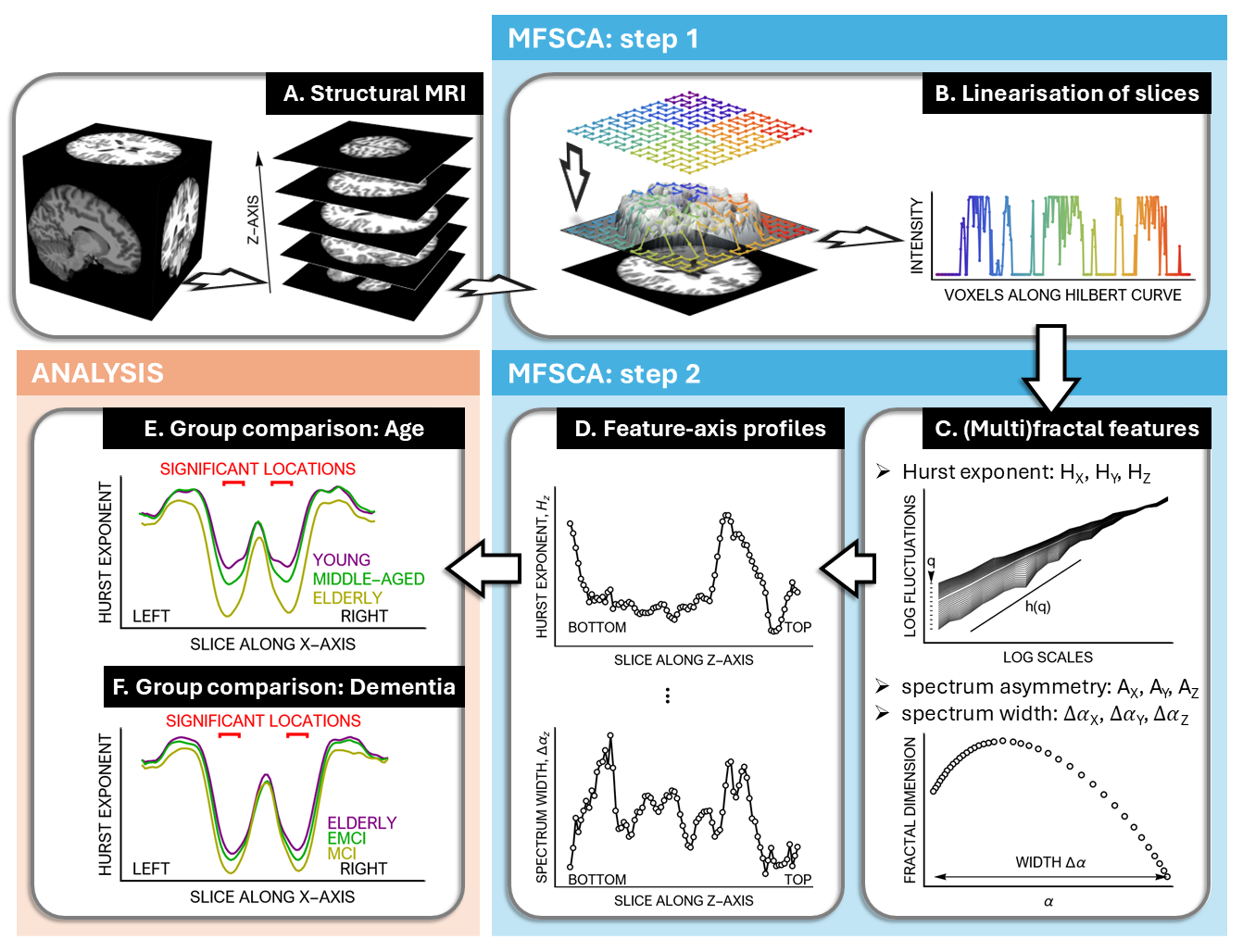}
\caption{\textbf{Flowchart of the Multifractal Space-filling Curve Analysis (MFSCA).} \textbf{A.} Skull-stripped MRI volume is analyzed slice by slice along a given axis. \textbf{B.} Each slice is overlaid with a space-filling curve (Hilbert curve). A one-dimensional representation of the slice is formed by voxel intensity along the curve. \textbf{C.} Multifractal detrended fluctuation analysis, see eq. (\ref{eq:fluct_function}), provides the generalized Hurst exponent $h(q)$ and the multifractal spectrum $f(\alpha)$. The Hurst exponent $H=h(2)$, the width of the spectrum $\Delta\alpha$, and its asymmetry $A$ are later used as features describing the slice. \textbf{D.} The multifractal features of all slices along a given axis are collected and plotted as brain profiles. Profiles along \textit{x}, \textit{y}, and \textit{z} axes of Hurst exponent $H_x, H_y, H_z$, spectrum width $\Delta\alpha_x, \Delta\alpha_y, \Delta\alpha_z$, and spectrum asymmetry $A_x, A_y, A_z$ are computed for individual subjects. \textbf{E. and F.} Sample results obtained for neuroimaging data. Brain profiles undergo group analyses in age cohorts: Young, Middle-Aged and Elderly Controls, see Fig.~\ref{fig:Haging}-\ref{fig:Daging}; and in dementia severity cohorts: Elderly Controls, Early Mild Cognitive Impairment, and Mild Cognitive Impairment, see Fig.~\ref{fig:Hdementia}-\ref{fig:Ddementia}. Per-slice statistical hypothesis testing identifies slices that are significant for distinguishing between the groups.}
\label{fig:flowchart}
\end{figure}

The full characterization of the organization of complex data is a challenging task, as it must account for both the linear and nonlinear properties of the data. The latter, although often present in real-world datasets, are frequently overlooked by standard correlation-based methods. Moreover, in the case of multidimensional data, the characterization of nonlinear relationships becomes particularly difficult due to the complex structure of the data under analysis. This motivates us to propose a novel methodology, Multifractal Space-filling Curve Analysis (MFSCA), for quantifying both linear and nonlinear relationships in data. MFSCA represents a generalization of the previously proposed Fractal Space Curve Analysis (FSCA) \cite{grela2025using}, which is primarily sensitive to linear dependencies in the data. The scheme of MFSCA, including its basic steps and possible organization of results, is depicted in Fig. \ref{fig:flowchart}. The method is divided into two main stages. The first stage involves transforming high-dimensional data into a time series. The second stage applies multifractal methodology to quantify the nonlinear organization of the data. The individual steps are described in detail in Sections \ref{sec:SFC} and \ref{sec:MFDFA} below. The example illustrated in the flowchart is based on the results of applying the proposed method to neuroimaging data. We note, however, that the method of analysis is not restricted to neurological data, but may include multidimensional data of any source.

\subsection{MFSCA: step 1.}
\label{sec:SFC}

\begin{figure}[H]
\centering
\includegraphics[width=1.0\columnwidth]{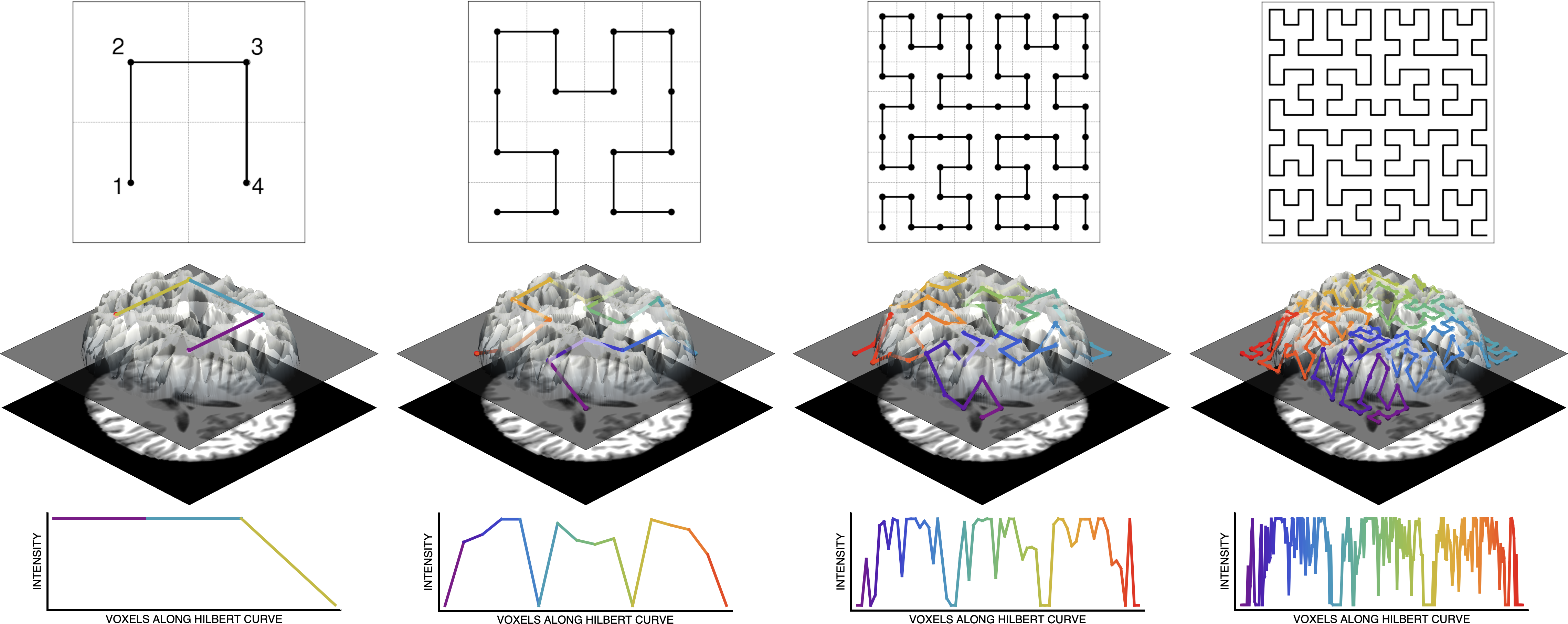}
\caption{Mapping voxel intensity into 1D. (Top) First four iterations of Hilbert space-filling curves on a square grid. In a given slice of MRI volume, the final iteration of the curve sets the order of traversal of all voxel centers of that slice. (Middle) Visualization of the MRI volume’s slice. The 3D landscape illustrates voxel intensities in that slice. The intensities are mapped onto the positions of the voxels along the curve. (Bottom) Linearized slice, i.e., one-dimensional representation of voxel intensities along the Hilbert curve. }
\label{fig:HilbertPeanoExamples}
\end{figure}

The first step in Multifractal Space-filling Curve Analysis (MFSCA) is to convert multidimensional data into a one-dimensional time series. To achieve this, we utilize the space-filling curve (SFC) technique, which effectively preserves the local properties of the multidimensional data within a one-dimensional framework~\cite{bader2012spacefilling}. The main idea is to order the data use the bijection mapping along a line that fills the multidimensional object. Various types of space-filling curves have been developed, with the Hilbert curve demonstrating particularly effective properties, as shown in reference \cite{grela2025using}. Figure \ref{fig:HilbertPeanoExamples} illustrates the construction of the Hilbert curve in the plane, and demonstrates how two-dimensional data can be transformed into a one-dimensional series. In addition, three-dimensional Hilbert curve constructions are available and can also be applied within the MFSFCA framework.

An important feature of the Hilbert curve is its Hölder continuity, which relates the distance along the curve, $|x-y|$, to the distance between their corresponding points in the embedded space $\parallel f(x) -f(y) \parallel$~\cite{bader2012spacefilling}:
\begin{equation}
\parallel f(x) -f(y) \parallel \leq c_f |x-y|^{1/d}
\label{eq:Holder_continuity}
\end{equation}
where $\parallel \cdot \parallel$ denotes the Euclidean norm, $d$ is a dimension of the curve, and $c_f$ is a constant.

\subsection{MFSCA: step 2.}
\label{sec:MFDFA}
To perform the multifractal analysis, the multifractal detrended fluctuation analysis (MFDFA) \cite{kantelhardt2002multifractal} is applied, which is one of the most effective methods for quantifying the multifractal properties of time series.

The MFDFA consists of the following steps. In the first step, the profile of the time series ${x(i)}$ $(i=1 \dots N)$ is calculated:
\begin{equation}\label{eq:1}
X(j)=\sum_{i=1}^{j} (x(i)-\langle x\rangle),\quad j=1,\dots,N,
\end{equation}
where $\langle\dots\rangle$ stands for averaging over the entire time series.

The profile is divided into $N_s$ non-overlapping segments labeled by $\nu$, each of length $s$, starting from the beginning of the time series. To account for the endpoint of the time series - especially when the total length of the series is not an integer multiple of $s$ - the division process is also repeated from the end of the time series. As a result, we ultimately obtain $2N_s$ segments. For each segment, the trend $P^{(m)}_\nu$ represented by the polynomial of degree $m$ fitted to the data is subtracted from the data, and the detrended variance is calculated according to the formula:
\begin{equation}
F^2(\nu,s)=\frac{1}{s}\sum_{j=1}^s\left\{X((\nu-1)s+j)-P^{(m)}_\nu(j)\right\}^2
\end{equation}
for segments $\nu=1,\dots,N_{s}$ and
\begin{equation}
F^2(\nu,s)=\frac{1}{s}\sum_{j=1}^s\left\{X(N-(\nu-N_s)s+j)-P^{(m)}_\nu(j)\right\}^2
\end{equation}
for segments $ N_{s}+1,\dots,2N_{s}$.
To quantify the scaling properties of the data in relation to the data amplitudes, the $q$-order fluctuations function is calculated according to the formula:
\begin{equation}
F_q(s)=\left(\frac{1}{2N_s}\sum_{\nu=1}^{2N_s}\left[F^2(\nu,s)\right]^{q/2}\right)^{1/q},\quad q\in\mathbb{R}\backslash{\{0\}}.
\label{eq:fluct_function}
\end{equation}
The parameter $q$ acts as a filter; therefore, for positive $q$, larger fluctuations are enhanced, while for negative $q$, smaller fluctuations receive more emphasis in the equation. In the case of fractal time series, the scaling relation is observed:
\begin{equation}
F_q(s)\sim s^{h(q)},
\end{equation}
where $h(q)$ represents the generalized Hurst exponent. For monofractal structures, the independence of $h$ from $q$ is observed such that $h(q) = H$. Conversely, for multifractal time series, $h(q)$ is a decreasing function. In the latter case, the Hurst exponent is obtained for $q = 2$, i.e., $h(2) = H$.

The generalized Hurst exponents further define the multifractal spectrum (singularity spectrum of H\"{o}lder exponents), $f(\alpha)$ \cite{kantelhardt2002multifractal}:
\begin{equation}
\alpha=h(q)+qh^\prime(q),\quad f(\alpha)=q[\alpha-h(q)]+1,
\label{eq:spectrum}
\end{equation}
where $h^\prime$ denotes a derivative of $h$, $\alpha$ determines the strength of the singularities, and $f(\alpha)$ can be viewed as the fractal dimension of a subset of the time series with singularities of magnitude $\alpha$. Usually, positively linearly correlated signals have their spectrum shifted towards $\alpha(q=2)>0.5$, and vice versa: negative linear autocorrelations shift the spectrum towards $\alpha(q=2)<0.5$. The maximum spectrum at $\alpha(q=2)=0.5$ suggests weak linear correlations or a lack thereof. Finally, the numerical features that we use to describe the multifractal spectrum: its width $\Delta\alpha$ and asymmetry $A$,  are computed as follows
\begin{equation}
\Delta\alpha = \alpha_{\max}-\alpha_{\min}, \quad
A = \frac{\alpha_0 - \alpha_{\min} - (\alpha_{\max} - \alpha_0)}{\Delta\alpha} = \frac{2\alpha_0 - \alpha_{\max} - \alpha_{\min}}{\Delta\alpha}
\label{eq:asymmetry}
\end{equation}
where $\alpha_{\min} = \min_q\alpha(q),\quad \alpha_{\max} = \max_q\alpha(q), \quad \alpha_0 =\text{argmax}_{\alpha} f(\alpha)$. The wider the multifractal spectrum (i.e., the larger the $\Delta \alpha $), the more complex the correlation structure of the data and the stronger the nonlinearities present in the system. Conversely, the smaller the $\Delta \alpha$, the weaker the nonlinearities contained in the data structure. The asymmetry coefficient indicates the organization of small and large fluctuations within the series. A negative value of $A$ indicates that the correlation structure is dominated by the organization of small fluctuations, whereas a positive value of $A$ indicates that the correlation structure is dominated by relationships among large fluctuations. $A = 0$ indicates a balance between the organisation of small and large fluctuations.

\section{Methods}
\label{methods-description}
\subsection{Dataset}

Data were drawn from the Open Access Series of Imaging Studies (OASIS) database. 
The OASIS-1~\cite{OASIS1} database comprises a cross-sectional collection of T1-weighted MRI scans of 416 subjects, aged 18-96 years, all right-handed. 
Of the participants older than 60 years, 100 were diagnosed with very mild to mild Alzheimer's disease. 
Accompanying clinical information includes Clinical Dementia Rating (CDR) and Mini-Mental State Examination (MMSE) scores. 
CDR is a dementia staging measure with a score of 0 indicating no symptoms, 0.5 -- very mild dementia, 1 -- mild dementia, and 2 -- moderate dementia. 
The MMSE is a brief, non-diagnostic questionnaire that assesses cognitive functioning, including recall, orientation, and language ability. 

Demographic data of the subjects is provided, including gender, age, handedness, education, and socioeconomic status (SES).  
Education was scored from 1 to 5, with 1 -- did not graduate high school, 2 -- graduated high school, 3 -- attended but did not graduate college, 4 -- graduated from college, 5 -- beyond college~\cite{OASIS1}. 
Socioeconomic status was assessed by the Hollingshead Index of Social Position, which is based on the weighted sum of occupation and education scores~\cite{wustl_isp_2000}. Occupation is scored for the primary earner in the participant's household, while education is recorded for the participant. The occupational status of retired participants is assessed based on the occupation before retirement.

For this analysis, the OASIS-1 subjects were organized into the following 5 cohorts according to age and CDR:
Young Control, Middle-Aged Control, Elderly Control, Early Mild Cognitive Impairment (EMCI), and Mild Cognitive Impairment (MCI). 
Inclusion criteria and cohort characteristics are provided in Table~\ref{tab:cohort-defs}, together with group sizes, mean age, and sex distribution.
Note that two healthy participants did not meet the inclusion criteria for any cohort and were excluded from the analysis. 
The cohort assignment creates three groups of elderly subjects with approximately equal average ages, enabling estimation of age- and disease-related associations. 

\subsection{Propensity Score Matching}

Ideally, the goal is to estimate causal effects of age and disease progression between cohorts on (multi)fractal characteristics of interest -- or, in the parlance of causal inference literature, effects of \emph{treatment} on an {outcome}. However, OASIS-1 is an observational study, so different co5horts also differ systematically in demographic characteristics. Such differences could bias our estimates of the treatment effect. 





To reduce this bias, we conduct cohort comparisons on weighted pseudo-populations of subjects, with weights assigned to subjects to ensure that demographic characteristics (covariates) are similarly distributed across cohorts.
Intuitively, this is achieved by assigning larger weights to subjects unlikely to belong to a cohort based on their demographic characteristics, and smaller weights otherwise. 
In other words, weighting is based on subjects' \emph{propensity} to belong to a cohort -- or, in the vocabulary of causal inference literature, their propensity to receive a certain \emph{treatment}. 
The approach is thus formally known as \emph{inverse probability of treatment weighting} (IPTW). The following limitations and assumptions of this technique guided our analytical choices and the interpretation of the results. 

Firstly, for the weighting to adequately balance the covariate distribution between cohorts, a subpopulation defined by a specific combination of covariate values must include representatives from all cohorts. This is the \emph{overlap (positivity)} assumption.
Secondly, variables included in the model as covariates should be associated with both treatment and outcome, but they must not lie on the causal pathway between treatment and outcome.
Moreover, estimating a truly causal effect would require measuring all confounder covariates. Certainly, some relevant covariates were not measured, such as cardiovascular risk factors (history of smoking, BMI).

Thus, when estimating the effect of disease progression, we included age and gender as covariates in the weighting model; when estimating the effect of aging, we included only gender. 
We excluded education because it was not assessed at all for young subjects (who did not complete their education), and it is missing for a substantial proportion of elderly subjects. 
We also excluded SES from the weighting model for the following reasons: (i) Subjects of low SES are poorly represented in OASIS-1, leading to violation of the positivity assumption; (ii) SES tends to increase over the course of a career, making it a potential mediator in the ageing analysis; (iii) SES may plausibly be affected by cognitive decline, making it a potential mediator in the disease progression analysis. 






\begin{table}[h]
\centering
\begin{tabular}{|l|ll|ll|l|}
\hline
\multicolumn{1}{|c|}{\multirow{2}{*}{Cohort}} &
\multicolumn{2}{c|}{Inclusion criteria} &
\multicolumn{3}{c|}{Characteristics} \\ \cline{2-6}
& \multicolumn{1}{l|}{Age} & CDR
& \multicolumn{1}{l|}{\# of subjects} & Mean age & \% male \\ \hline

Young Control
& \multicolumn{1}{l|}{age\textless{}40} & not assessed
& \multicolumn{1}{l|}{152} & 23.2 (4.2) & 46.1  \\ \hline

Middle Aged Control
& \multicolumn{1}{l|}{40$\leq$age$<$60} &
\begin{tabular}[c]{@{}l@{}}not assessed\\ or CDR=0\end{tabular}
& \multicolumn{1}{l|}{64} & 50.1 (5.3) &  32.8 \\ \hline 

Elderly Control
& \multicolumn{1}{l|}{age$\geq$60} & CDR=0
& \multicolumn{1}{l|}{98} & 75.9 (9.0) & 26.5  \\ \hline

EMCI
& \multicolumn{1}{l|}{} & CDR=0.5
& \multicolumn{1}{l|}{70} & 76.2 (7.2) &  44.3 \\ \hline

MCI
& \multicolumn{1}{l|}{} & CDR>0.5
& \multicolumn{1}{l|}{30} & 78.0 (6.9) & 33.3 \\ \hline

\end{tabular}
\caption{OASIS-1 cohort definitions and basic characteristics. \textit{Abbreviations:} CDR -- Clinical Dementia Rating; EMCI -- Early Mild Cognitive Impairment; MCI -- Mild Cognitive Impairment.}
\label{tab:cohort-defs}
\end{table}

\subsection{Linear Discriminant Analysis} \label{sec:methodsLDA}

In order to explore whether the fractal and multifractal features provide informative characterization of the cohorts several dimensionality reduction techniques were tested. They include standard unsupervised methods such as Principal Component Analysis (PCA), Uniform Manifold Approximation and Projection (UMAP), t-distributed Stochastic Neighbor Embedding (t-SNE)~\cite{vandermaaten2008tsne} and several versions of Linear Discriminant Analysis (LDA), which is a supervised technique.
LDA was specifically chosen, because it is designed to reduce dimensionality while maximizing separability of the data assigned different class labels (in our case, the cohorts). The separability is defined as the ratio of the distance between the cohort centroids and the within-cohort variance.
As a result, this method can indicate which features were important for the discrimination between cohorts.

In the final stage of the analysis there are in total 414 subject (cf. Table~\ref{tab:cohort-defs}) and 777 features (each metric $H$, $\Delta\alpha$, $A$ for x, y and z axes with 94, 81 and 84 accepted slices, respectively) which results in the small sample size problem: (a) having more features than data points can easily lead to overfitting, (b) the within-class scatter matrix (which encodes the said within-cohort variance) is not invertible.
For this reason, the PCA-LDA pipeline was performed with 50 most important principal components (reducing dimensionality from 777 features) and Fisher's LDA reducing it further to 2 linear discriminants.
The chosen LDA version involves the maximization criterion (see~\cite{theodoridis2009Pattern} for alternatives):
$$
J_3 = \mathrm{trace}(S^{+}_w S_b)
$$
where $S_w$ is the within-class scatter matrix, $S_b$ the between-class scatter matrix, and $S^{+}$ is the pseudoinverse.
No additional shrinkage of $S_w$ was performed, because the PCA step effectively serves also as a regularizer. The trade-off is, however, that during the dimension reduction part of the discriminatory information may be lost.

\section{Results}
\label{Results}
\subsection{Deterministic Cascades} 
\label{sec:cascades}

To assess the sensitivity of MFSCA to nonlinear correlations in data, we tested our method on artificially generated multifractal structures. For this purpose, we used two-dimensional multifractal cascades, well-known multiscale structures with theoretically derivable properties. As we mentioned before, the crucial multifractal characteristic is the multifractal spectrum, which quantifies the data's complexity. Namely, the broader the spectrum, the more developed the multifractality is, and the greater the degree of nonlinear correlations.  For deterministic cascades, the following equations give the analytical expressions for the multifractal spectrum \cite{Cheng2005,Xu2017}: 
\begin{equation}
\left\{
\begin{alignedat}{2}
\alpha(q) &= 
\frac{p_1^q \ln p_1 + p_2^q \ln p_2 + p_3^q \ln p_3 + p_4^q \ln p_4}
{(p_1^q + p_2^q + p_3^q + p_4^q)\ln 2}, \\
f(\alpha) &= 
\frac{p_1 \ln p_1 + p_2 \ln p_2 + p_3 \ln p_3 + p_4 \ln p_4}{\ln 2}.
\end{alignedat}
\right.
\end{equation}
where $p_x$ denotes the weights used to generate the two-dimensional multifractal measure. 
We generated a set of multifractal cascades with varying degrees of multifractality, spanning a spectrum width of [1.5, 2] with a step size of 0.05. 

\begin{figure}[htbp]
    \includegraphics[scale=0.15]{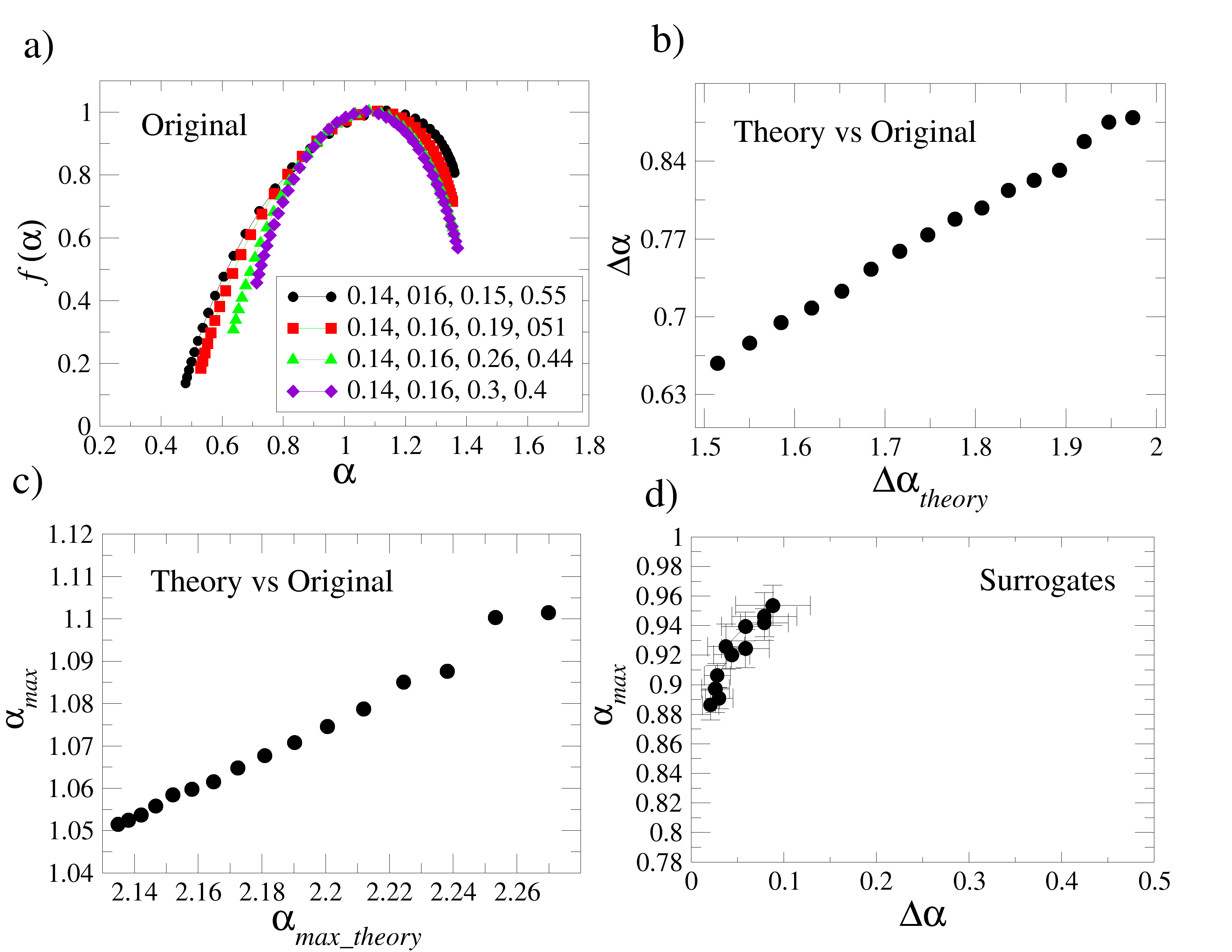}
    \caption{MFSCA analysis of the deterministic cascades. a) Multifractal spectra obtained for the sample cascades with different degrees of complexity (width of the multifractal spectrum). b) Comparison of the estimated widths of the multifractal spectrum with its theoretical counterparts. c) comparison of the maxima localisation estimated spectra $f(\alpha)$ with theoretical counterparts. d) multifractal spectrum characteristics for the surrogate data - only linear correlations are preserved. Error bars denote the standard deviation over 10 realizations.}
    \label{fig:cascade}
\end{figure}

The results of this analysis are shown in Fig. \ref{fig:cascade}. First, we compare the position of the spectrum's maximum with its theoretical value (Fig. \ref{fig:cascade}c). The linear relationship between these quantities, with an estimated slope coefficient of 0.5, aligns perfectly with the theoretical relation (see Eq.~\ref{eq:Holder_continuity}). A similar relationship is observed for the spectrum width (Fig. \ref{fig:cascade}b): in this case as well, the proportional coefficient between the empirical and theoretical results is approximately 0.5. These findings confirm that the methodology is valid and that the nonlinearity present in the data is accurately quantified. 

As the next test, we generated surrogate data by removing nonlinear correlations while preserving linear ones. To achieve this, we employed a Fourier transform (FT)-based procedure. The empirical data were transformed from the spatial domain to the frequency domain, in which the phases of the FT were randomly shuffled while the amplitudes---which determine the power spectrum, which quantifies the linear dependence of the data---were preserved. By applying the inverse FT, we return the data to the original domain, resulting in a time series that retains only linear correlations, thus only possible monofractal organization. 

In this analysis, we applied the two-dimensional Fourier transform to generate surrogates of the multifractal cascades. For each cascade, we produced ten surrogate datasets. Then we performed MFSCA on these surrogates, and the results are shown in Fig.~\ref{fig:cascade}d. Notably, the maxima of the estimated spectra remained unchanged after the surrogate procedure; however, the spectral widths narrowed significantly for the surrogates, indicating monofractality. Thus, MFSCA accurately detected both the linear and nonlinear spatial organization of the data, demonstrating high sensitivity to multifractal structure.

\subsection{Multifractal Analysis of MRI Data in Alzheimer’s Disease Progression}
\label{sec:results_AD}

\subsubsection{MFSCA Results for Representative Cases}

For the MRI images obtained from each subject, we analyzed volume slices along the axial, sagittal, and coronal directions, denoted in the paper as $x$, $y$, and $z$, respectively. The MFSCA procedure was applied to each image, linearizing each slice using the Hilbert curve. For each slice, the resulting one-dimensional series was then analyzed using MFDFA to estimate the fluctuation functions and the multifractal spectra.

\begin{figure}[htbp]
    \includegraphics[scale=0.12]{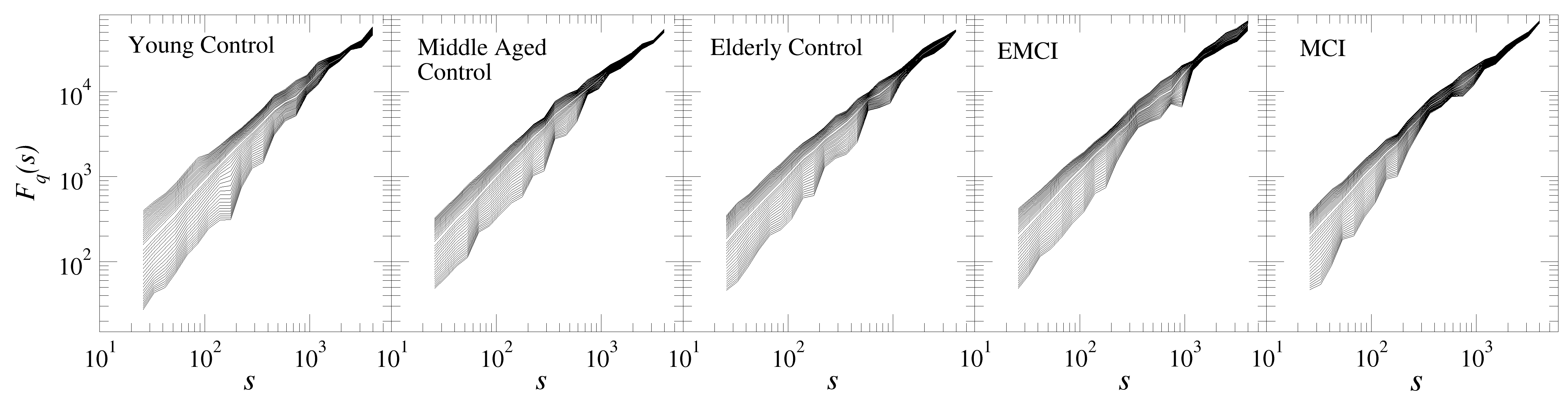}
    \includegraphics[scale=0.10]{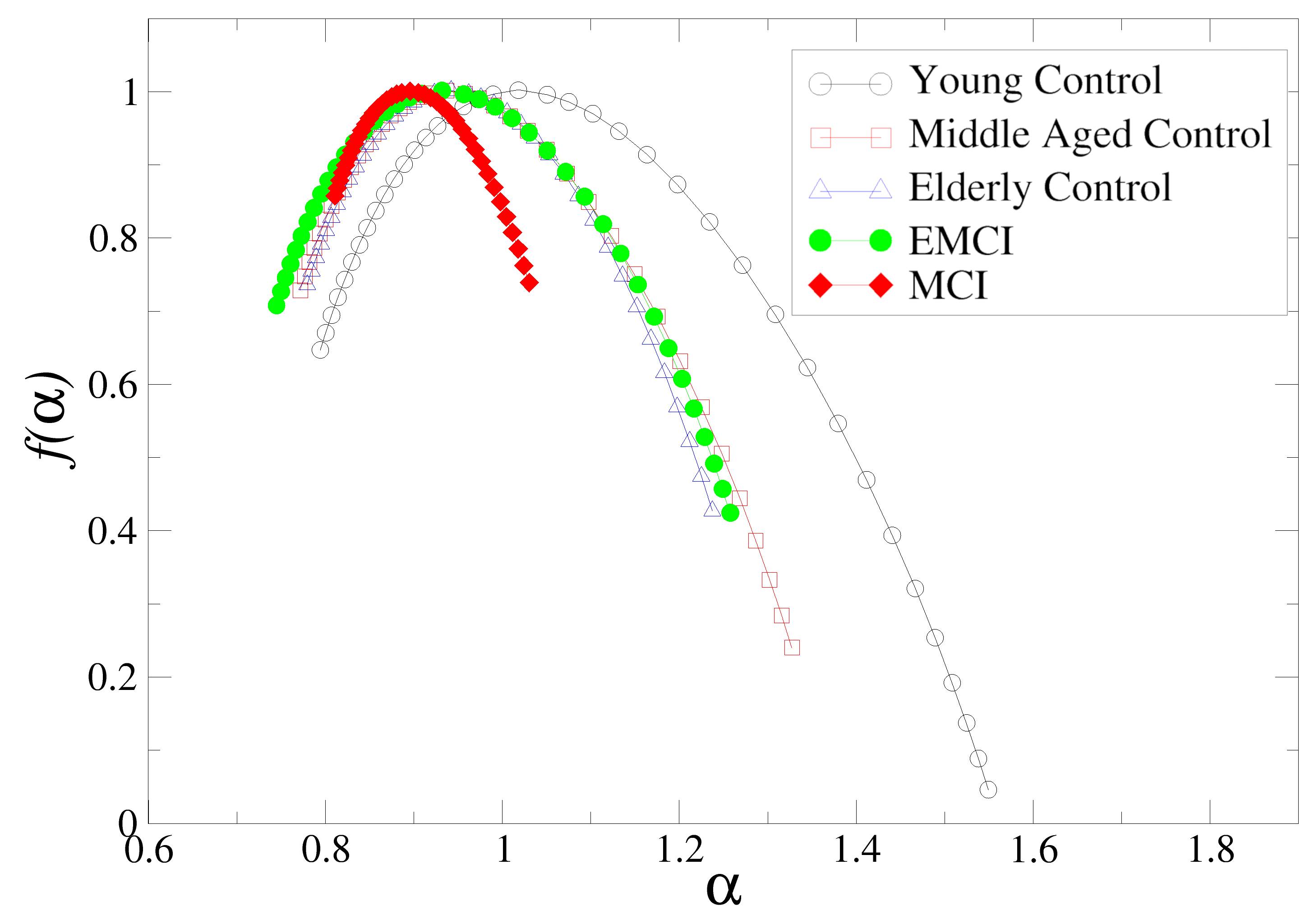}
    \caption{(Top) Fluctuation functions for slice 129 along the y-axis (sagittal) for representative individuals from each cohort. (Bottom) Corresponding multifractal spectra.}
    \label{fig:fluctuatio_funtion_sample}
\end{figure}

The examples of fluctuation functions for each cohort are shown in Fig. \ref{fig:fluctuatio_funtion_sample}. Already a visual inspection reveals a clear multiscale organization of the data in all groups, reflected in the heterogeneous scaling behavior of $F_q(s)$ and the family of slopes that vary with the moment $q$. However, the quality of scaling depends strongly on the cohort. The most pronounced power-law behavior is observed in the Young Control group, where the scaling range extends over nearly two orders of magnitude. In contrast, the strongest deviations from ideal scaling occur in the EMCI and MCI cohorts, for which the scaling of $F_q(s)$ for different $q$ collapses toward monoscaling at the largest scales.

The multifractal spectra derived from the $F_q(s)$ functions quantitatively characterize these differences in scaling behavior in the bottom panel of Fig.~\ref{fig:fluctuatio_funtion_sample}. The widest spectra are found in the Young Control group, confirming the presence of a rich multifractal structure. The right-skewed shape of the spectrum suggests that this multifractality arises from the heterogeneous organization of subtle grayscale variations in normal brain anatomy. In the remaining groups, the spectra become progressively narrower, which may reflect the combined effects of aging (Middle-Aged Control and Elderly Control) and disease progression (EMCI and MCI). The narrowest spectrum is observed in the MCI group, indicating markedly reduced multifractal behavior.

\subsubsection{MFSCA Analysis Across Groups with Different Dementia Progression Stages}
\label{sec:demenatia}

Our analyses were divided into two parts. First, we estimated multifractal profiles for the control group across different age categories, and then compared them with those obtained for the subject at different stages of dementia. 

\begin{figure}[ht]
    \centering
    \includegraphics[width=\linewidth]{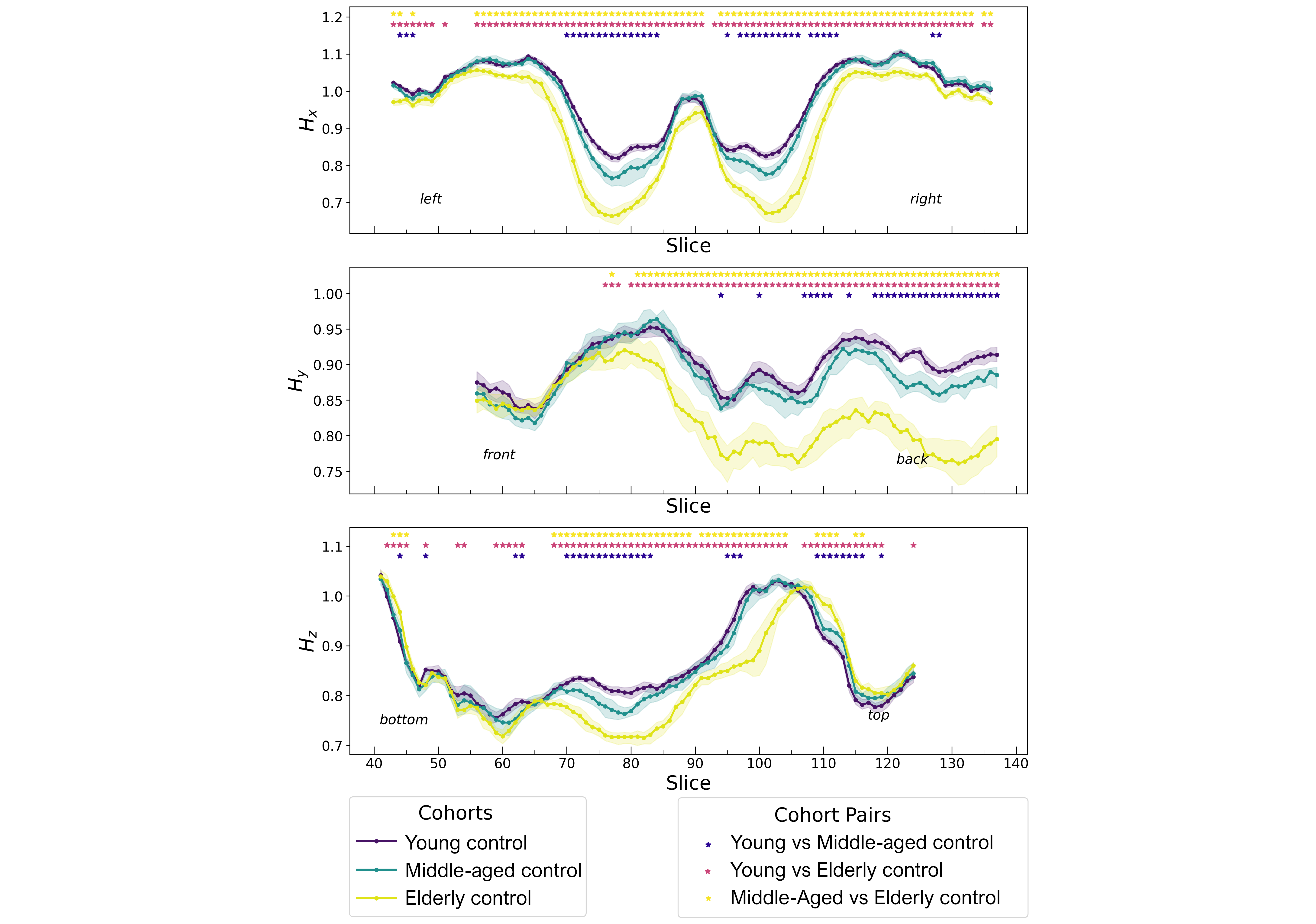}
    \caption{Slice-wise profiles of the long-scale Hurst exponent $H$ for the Young, Middle-Aged, and Elderly Control cohorts in the OASIS-1 dataset. 
    The panels show profiles along the $x$ (top), $y$ (middle), $z$ (bottom) axes. In each panel, the cohort median H value at each slice is indicated by a dot, and neighbouring points are connected to highlight the profile shape. 
    The shaded bands indicate the 95\% bootstrap confidence interval of the median ($n_{boot} = 5000$). 
    The labels inside the panels indicate the anatomical orientation of the slice index. 
    The stars mark slices at which the corresponding pair of cohorts differs significantly, based on a bootstrap test with within-slice Holm correction ($p < 0.05$). 
    Older cohorts tend to show lower $H$ values.}
    \label{fig:Haging}
\end{figure}

\begin{figure}[ht]
    \centering
    \includegraphics[width=\linewidth]{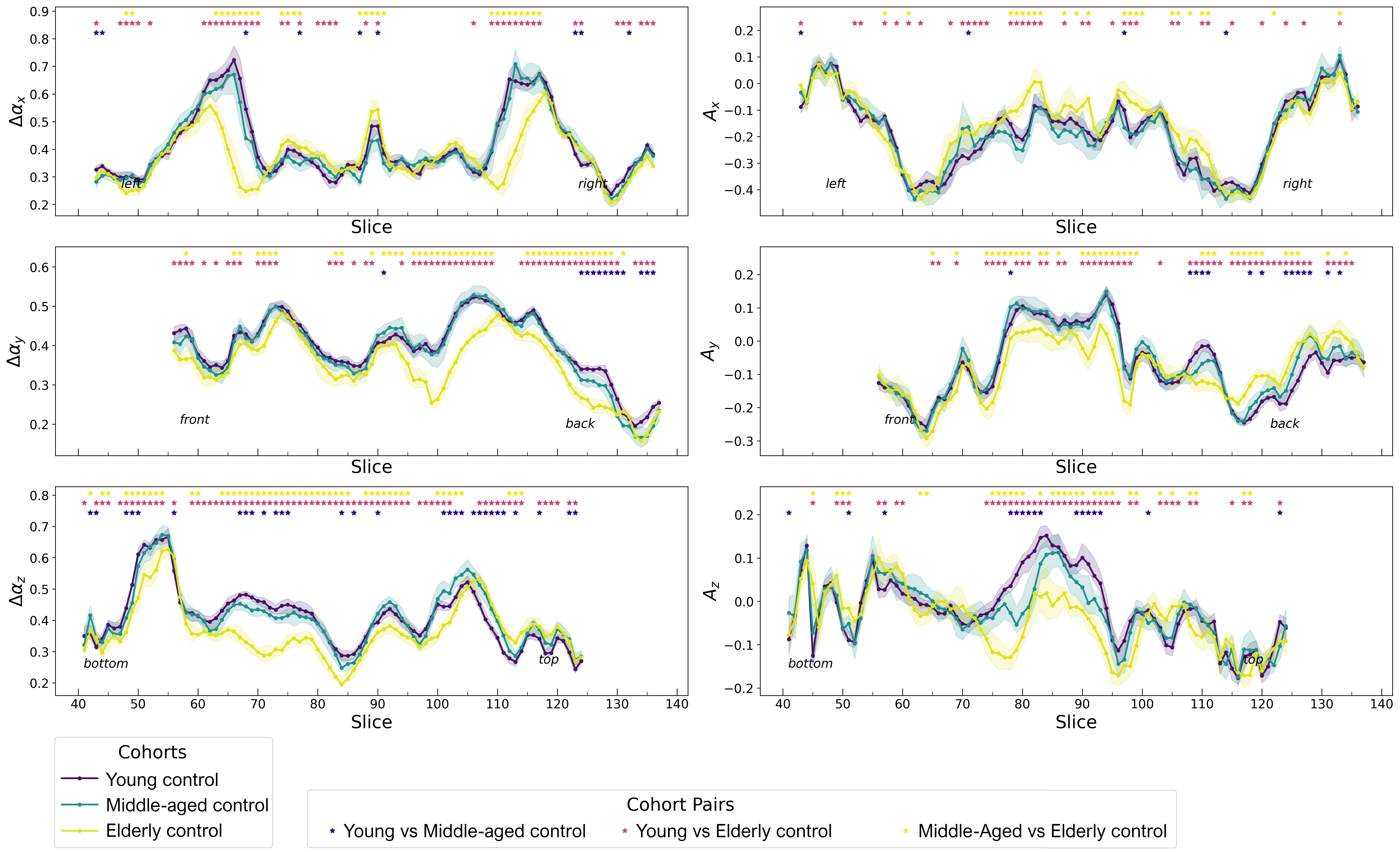}  
    \caption{Slice-wise profiles of the width $\Delta\alpha$ (left column) and asymmetry (right column) of the multifractal spectrum for the Young, Middle-Aged, and Elderly Control cohorts.
    The rows show profiles along the x (top), y (middle), and z (bottom) axes. All markers follow the same conventions as in  Fig.~\ref{fig:Haging}.}
    \label{fig:Daging}
\end{figure}

To obtain a comprehensive view of the relationship between multifractal characteristics and age-related development in healthy subjects in the control group, we analyzed the results by age. We considered three age groups: under 40 years (Young Control), between 40 and 60 years (Middle-Aged Control), and above 60 years (Elderly Control) (cf. Table \ref{tab:cohort-defs}). For each subject, we estimated multifractal characteristics for each image slice and for each direction (x, y, z). We then computed the group multifractal profile by averaging the results across all slices of each subject. The aggregated results are presented in Figure \ref{fig:Haging} and \ref{fig:Daging}.

As we can see, the Hurst exponent, the width of the multifractal spectra, and spectrum asymmetry vary across brain regions; however, the differences in profiles between groups are readily apparent. For the Young Control group, both the Hurst exponent and the width of the multifractal spectrum reach their highest values, indicating a well-developed correlation structure at both linear and nonlinear levels in brain organization, as reflected in the images. The spectrum asymmetry takes negative or near-zero values (for the x-axis) and positive values (for the y- and z-axes), suggesting that the spatial organization of nonlinear correlations influences both small and large variations in brain structure, depending on the direction considered. Comparable results were observed for the Middle-Aged Control group, with only slightly lower values in specific brain regions.

However, in the Elderly Control group, the multifractal characteristics indicate a substantially less-developed correlation structure. In particular, the values of the Hurst exponent, the multifractal spectrum width, and the spectrum asymmetry are significantly lower across a range of scans. The reduction in multifractal characteristics across all three spatial directions suggests a shift toward a more monofractal structure, characterised by weaker nonlinear organisation and a closer resemblance to white-noise processes. 

\begin{figure}[ht]
    \includegraphics[scale=0.3]{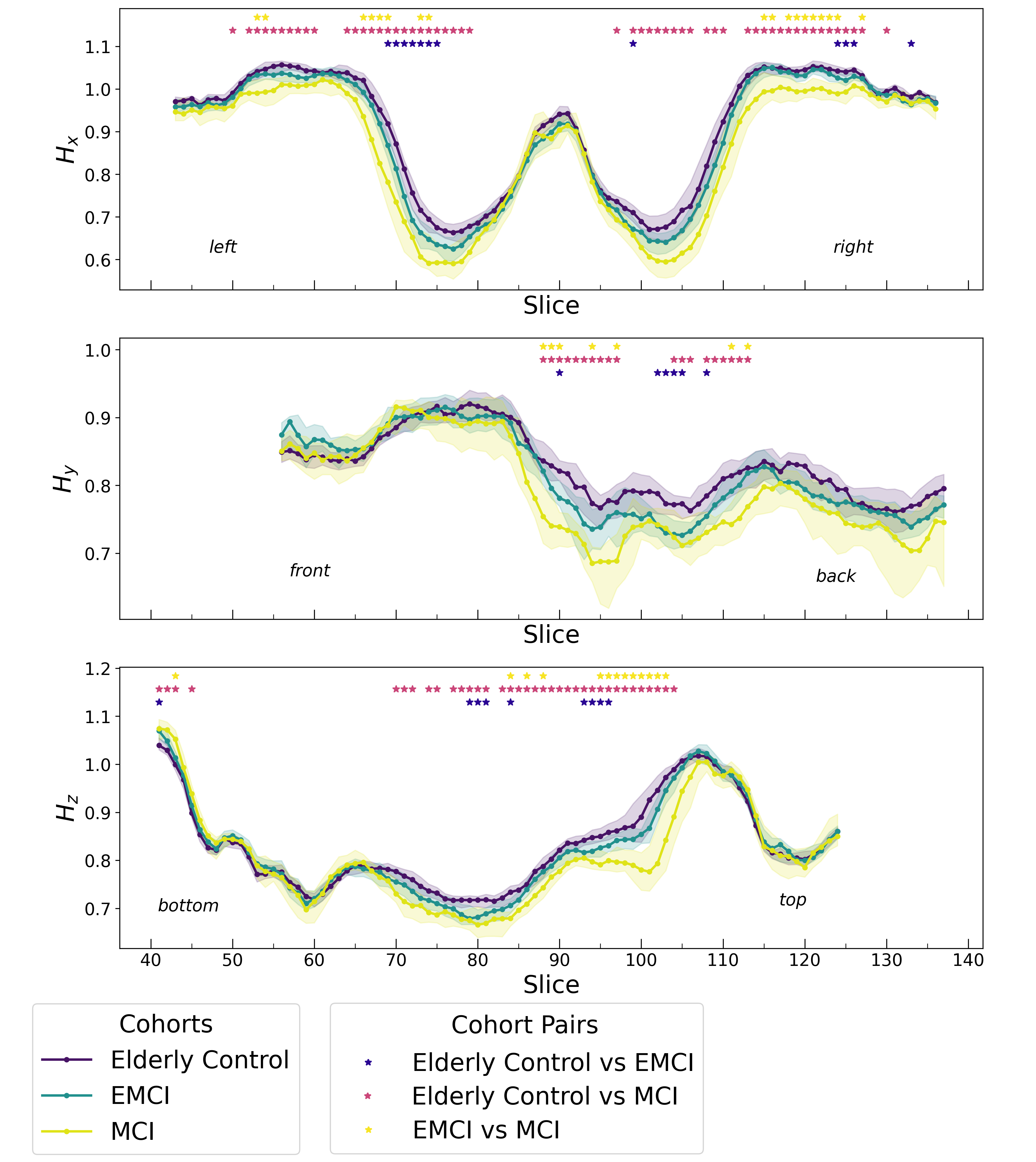}
    \caption{Slice-wise profiles of the long-scale Hurst exponent $H$ for the Elderly Control, EMCI, and MCI cohorts. 
    All markers follow the same conventions as in  Figs~\ref{fig:Haging}-\ref{fig:Daging}. 
    More cognitively impaired cohorts tend to show lower $H$ values.
    }
    \label{fig:Hdementia}
\end{figure}

\begin{figure}[ht]
    \centering
    \includegraphics[width=\linewidth]{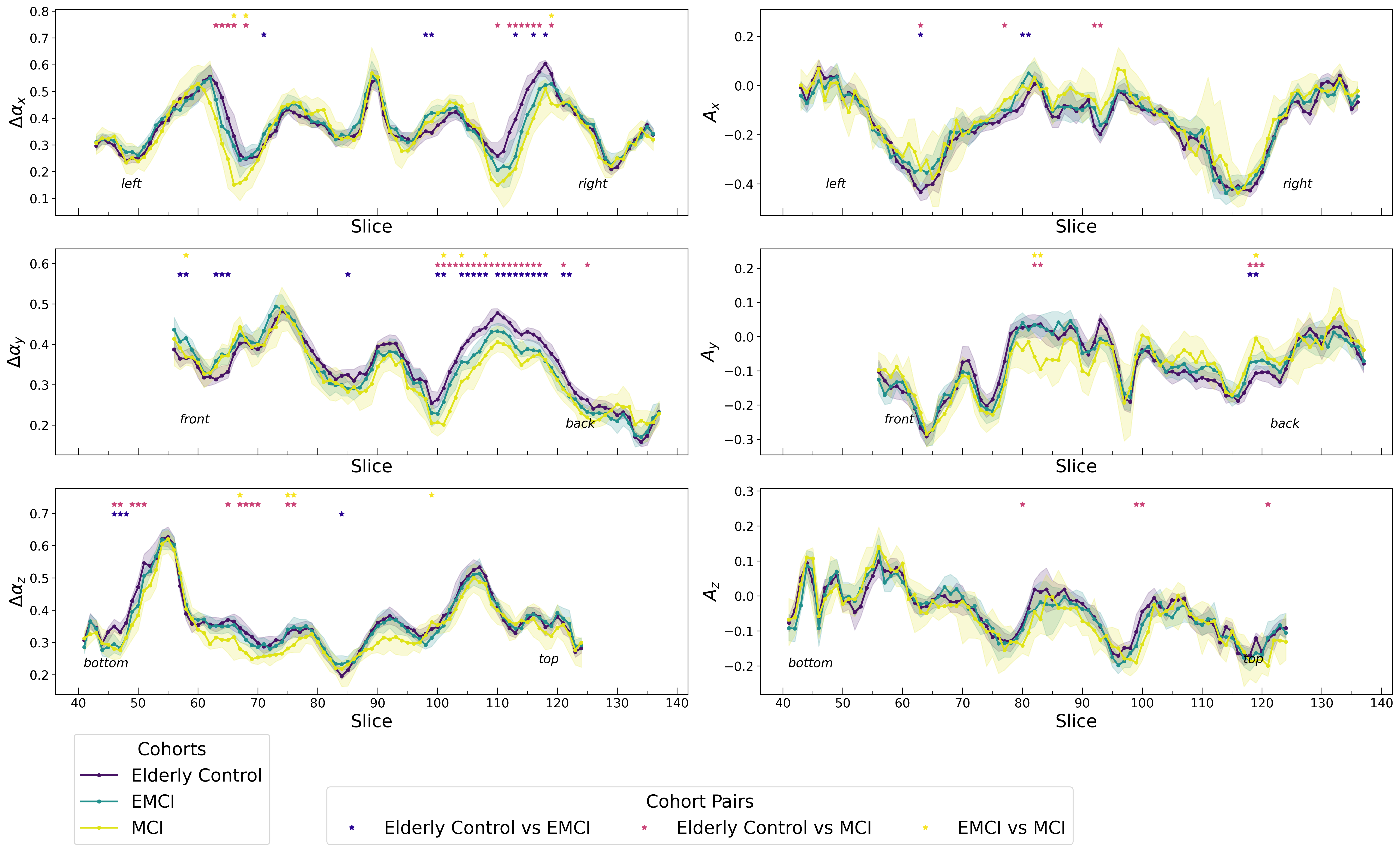} 
    \caption{Slice-wise profiles of width $\Delta\alpha$ (left column) and asymmetry (right column) of the multifractal spectrum for the Elderly Control, EMCI, and MCI cohorts in the OASIS-1 dataset.
    The rows show profiles along the x (top), y (middle), and z (bottom) axes.
    All markers follow the same conventions as in  Figs~\ref{fig:Haging}-\ref{fig:Daging}. }
    \label{fig:Ddementia}
\end{figure}

In the next stage, we compare the Elderly Control group with dementia patients, as these groups are best matched in terms of age. We include both EMCI and MCI groups in the analysis. As in the previous case, we calculate the Hurst exponent, the width of the multifractal spectrum, and its asymmetry as indicators of data organisation. These quantities are then averaged over the participants within the considered slices to obtain the multifractal profiles. The results of these calculations are presented in Figures \ref{fig:Hdementia} and \ref{fig:Ddementia}.

\begin{figure}[ht]
    \centering
    \includegraphics[width=\linewidth]{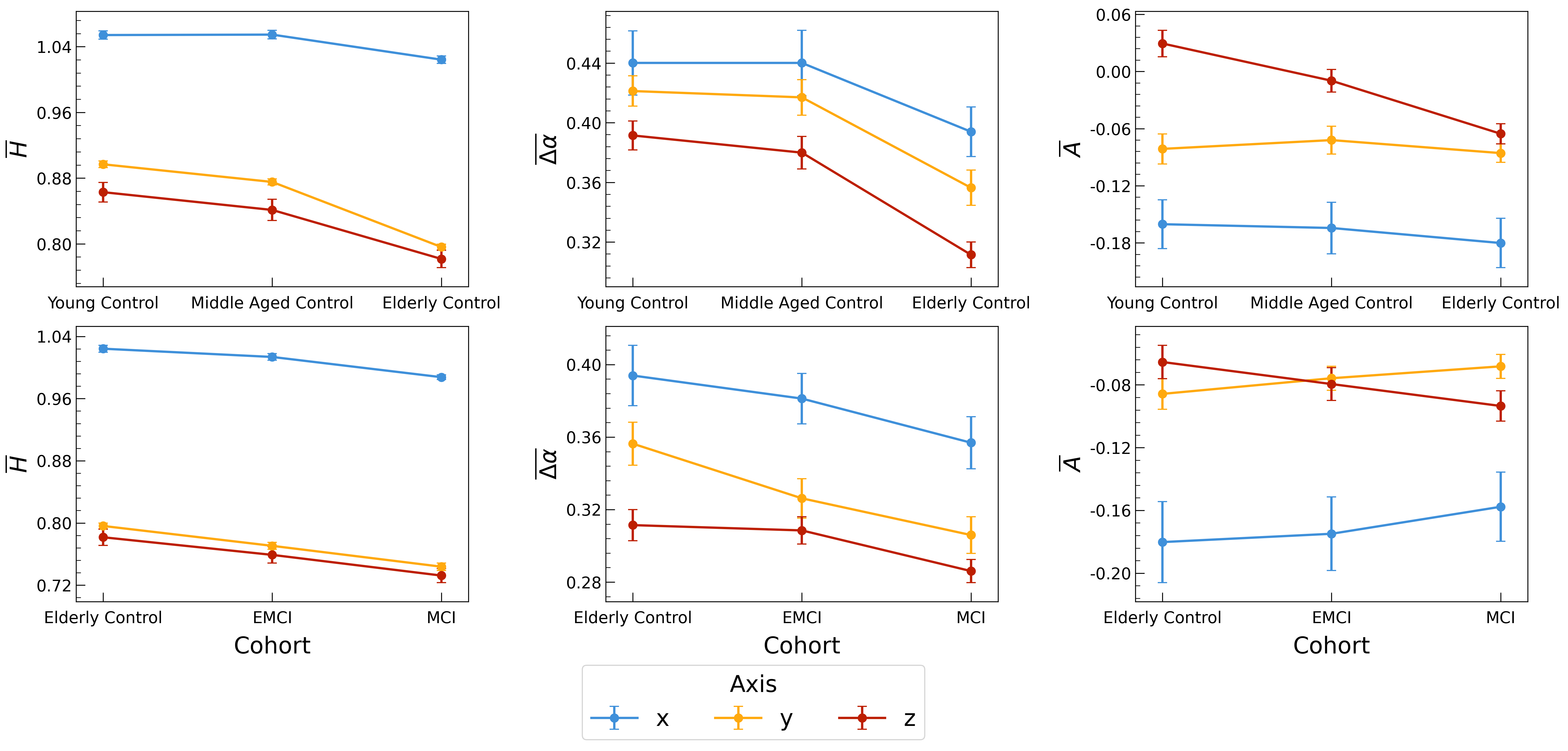} 
    \caption{Slice profile averages (Top) for the Young, Middle-Aged, and Elderly Control cohorts and (Bottom) for the Elderly Control, EMCI, and MCI cohorts computed within selected slice ranges. 
    The panels show the average of the slice median values of the long-scale Hurst exponent $\overline{H}$ (left), the width $\overline{\Delta\alpha}$ (middle), and asymmetry $\overline{A}$ (right) of the multifractal spectrum, separately for each axis. 
    Error bars denote the standard error of the mean. 
    The slice ranges were selected separately for each axis and correspond to regions where $H$ is relatively stable.}
    \label{fig:avg_aging}
\end{figure}

The changes in multifractal characteristics visibly follow a pattern similar to that observed for age-related effects.
See the bottom panel of Fig.~\ref{fig:avg_aging}) for trends in the whole brain averages (i.e., naive averages of the profiles in Figs. \ref{fig:Hdementia}-\ref{fig:Ddementia} over slice axis).
All indicators suggest that the multifractality of data from subjects with dementia is substantially reduced compared to the Elderly Control group. Furthermore, these changes in brain structure are associated with the stage of dementia progression. The highest degree of multifractality is observed in the Elderly Control group, whereas dementia leads to a progressive reduction in brain organization. This reduction is more pronounced in the EMCI and MCI groups, with the latter exhibiting behaviour closest to a monofractal structure.


\subsection{Identification of the Importance of Multifractal Parameters Using Dimensionality Reduction}

In this section we go beyond statistical hypothesis testing of group differences based on individual slices. We are rather interested in (i) exploring what data structure do all available (multi)fractal metrics reveal, (ii) group classification based on these metrics, (iii) assessing their importance for distinguishing a particular group.

\begin{figure}[ht]
    \centering
    \includegraphics[scale=0.42]{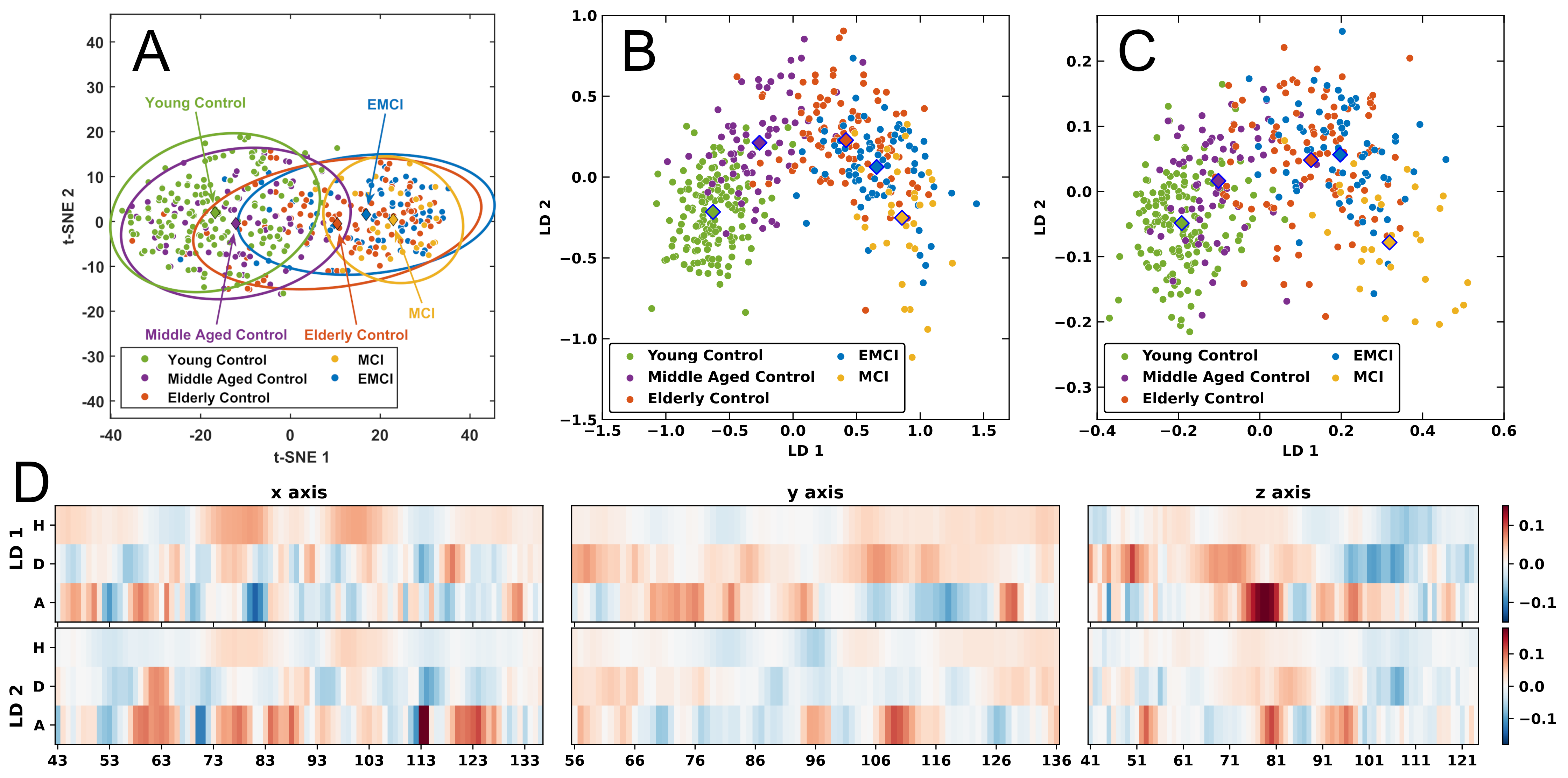}
    \caption{Dimensionality reduction of all features (all metrics: $H$, $A$ and $D$ on each slice of all three axes). (A) unsupervised dimensionality reduction with t-SNE. (B) supervised dimensionality reduction with LDA. (C) LDA only on multifractal features ($A$ and $\Delta\alpha$). (D) Linear discriminants from (B) indicating features contributing to the discrimination of all subject groups. In (A-C) the colors indicate the subject groups and the larger diamond markers are group means. In (A) the ellipses are group-specific 90\% confidence regions.}
    \label{fig:centroidy}
\end{figure}

Starting with (i), the general relationship between the examined cohorts is illustrated in Fig.~\ref{fig:centroidy} A, where t-SNE was employed to reduce the dimensionality from almost 800 features ($H$, $\Delta\alpha$, and $A$ metrics for each slice cut along x, y, and z axes) to a two-dimensional space. 
In that plot, points represent individual subjects, whereas the larger diamonds indicate the centroids of the respective groups in the embedded space. In addition, ellipses denote cohort-specific density regions estimated from the covariance structure and correspond to the 90\% confidence level.

The results show that the principal direction \textit{t-SNE 1} encodes both the age and severity of the cognitive impairment: the cohort means are ordered Young Control < Middle-Aged Control < Elderly Control < EMCI < MCI with the largest gap between the middle-aged and elderly and almost no overlap between the middle-aged and MCI. MCI is the most compact group, while the elderly and EMCI are the most dispersed along this direction, overlapping with all the other groups.
This exploratory technique alone does not allow to disentangle age from neurodegeneration, but rather locates neurodegeneration on the edge of a continuum of aging.

Moving to (ii), the LDA is a supervised (i.e. aware of the cohort labels) classification technique, which however also performs dimensional reduction. The linear discriminant space serves as a denoised feature space maximizing discrimination between the groups. In Fig.~\ref{fig:centroidy}, the projections represent reduction from all features in panel B, or only from the multifractal ones ($A$ and $\Delta\alpha$) in panel C. There is a visible order of the cohorts, similar to the one in t-SNE, which follows a U-shape in two dimensions. Again, the principal discriminant direction allows follows aging, while the second direction contains additional information that allows differentiate groups of similar age (the youth from the middle-aged, and the elderly from EMCI and MCI). Solely the multifractal features seem to contain enough information to perform that discrimination.

This method is also a way of obtaining (iii) feature importance, as presented in panel D of Fig.~\ref{fig:centroidy}. There the color intensity encodes how much each metric contributes to the discriminant directions LD 1 and LD 2.
These contributions roughly correspond to the metrics and slices where in Figures~\ref{fig:Hdementia}-\ref{fig:Daging} one can spot the largest differences between groups, e.g.:
$H$ around slices 70-80 and 95-105 along x-axis, $\Delta\alpha$ around slices 95-110 in y-axis, $A$ around slices 70-80 along z-axis, and so on.
In essence, this plot illustrates that both fractal and multifractal metrics can be informative for cohort discrimination.

By comparing feature contributions between cohort pairs it is further possible to indicate which metrics are associated with different stages of aging (young cohort vs. the middle-aged, and middle-aged vs. the elderly) and which with the disease progression (the elderly vs. EMCI and EMCI vs. MCI).
As a reference point, the raw features (without using LDA) in Supplementary Figure~\ref{supp_fig:raw_prototypes} reveal among others that $\Delta\alpha$ in slices around 65 and 115 along x-axis (and similarly $H$ around slice 100 on z-axis) progressively differs between the consecutive cohorts (although they are the strongest between the middle-aged and the elderly), which could be interpreted as neurodegeneration being an extension of the aging process.
On the other hand, when looking at the feature differences after PCA-LDA procedure in Supplementary Figure~\ref{supp_fig:LDA_prototypes}, the picture changes.
In general, the multifractal features strengthen the discrimination even though the differences in raw values are smaller in magnitude. The differences derived from PCA-LDA, however, are noisier (less smooth spatially between slices) in $\Delta\alpha$ and $A$, which might suggest that the model is still overfitting.
Examples of more detailed observations for the LDA differences are: (a) $H$ is mostly useful in discrimination between the youth, middle-aged, elderly, but not for cognitive impairment progression; (b) the features most strongly discriminating the middle-aged vs. elderly, but also the elderly vs. EMCI and EMCI vs. MCI, are asymmetry features (e.g., around slices 70-80 and 125-130 on y axis and slices around 70-80 on z axis); (c) $A_z$ difference seems to progresively migrate from slices around 80 to 70, as if the location of the change was age- and disease-related.

\section{Conclusions}
\label{sec:conclusions}


In this work, we propose a novel methodology for analyzing multidimensional data, particularly suited to nonlinear datasets that are difficult to capture using standard analytical approaches. In the first step, the method employs space-filling curve techniques to linearise multidimensional data, transforming them into a one-dimensional time series. To reduce dimensionality while preserving local structural properties, we use the Hilbert curve as a reliable mapping. In the second step, the resulting time series is analysed using Multifractal Detrended Fluctuation Analysis (MFDFA) to quantify its fractal characteristics. To demonstrate the effectiveness of the proposed methodology, we apply it to both artificially generated two-dimensional multifractal cascades with prescribed properties and real data obtained from patients with dementia. 

In the case of artificial data, the method successfully recovers both linear and nonlinear features, as quantified by the multifractal spectrum. By applying Fourier phase-randomized surrogates to the images, we generate two-dimensional data that preserve the same linear correlations while eliminating nonlinear structure. Our methodology accurately captures these differences in data organisation, underscoring the importance of testing for nonlinear relationships.

For the real data analysis, we examine data obtained from patients at two stages of disease progression: EMCI (Early Mild Cognitive Impairment) and MCI (Mild Cognitive Impairment). As a reference, we consider a healthy control group divided into three age categories: Young, Middle-aged, and Elderly (see Table \ref{tab:cohort-defs} for details).
As the MFSCA method generates hundreds of features corresponding to various brain locations, it opens the possibility of utilizing various statistical and machine learning methods for prediction and interpretation.

The statistial analysis yields several novel findings. First, the results for the control groups indicate that ageing significantly influences the scaling properties of brain organisation. In particular, we observe a systematic transition from multifractality toward monofractality (see Fig.~\ref{fig:avg_aging}), corresponding to a reduction in nonlinear characteristics with increasing age. Additionally, linear correlations weaken over time, causing the signal to become progressively closer to white noise.
This attenuation of heterogeneous scaling properties is even more pronounced in patients with dementia. When comparing age-matched groups, specifically, the Elderly controls and dementia patients, we observe a further reduction in multiscale brain organisation, which is systematically associated with disease progression.
The dimensionality reduction methods point to the conclusion that there is a common trend across brain areas and fractal metrics, that is jointly symptomatic of both aging and progression of dementia.
Other machine learning methods, however, are also able to extract subtler and more precisely characterized changes associated with disease progression only.

\section*{Acknowledgements}

The research for this publication has been supported by a grant from the Priority Research Area DigiWorld under the Strategic Programme Excellence Initiative at Jagiellonian University.

Data were provided [in part] by OASIS-1: Cross-Sectional: Principal Investigators: D. Marcus, R, Buckner, J, Csernansky J. Morris; P50 AG05681, P01 AG03991, P01 AG026276, R01 AG021910, P20 MH071616, U24 RR021382.




\section*{Data availability statement}
Data are available from the original sources:~\cite{ADNI,OASIS1,Poldrack2015}.
Code is available in \url{https://github.com/Mark-Kac-Center/MFSCA}.
Correspondence should be addressed to Jeremi Ochab (jeremi.ochab@uj.edu.pl).

\section*{Conflict of interests}
The authors declare no competing interests.


% ...


\section*{Supplement}
\subsection{OASIS-1}

\begin{figure}[ht]
    \centering
    \includegraphics[scale=0.4]{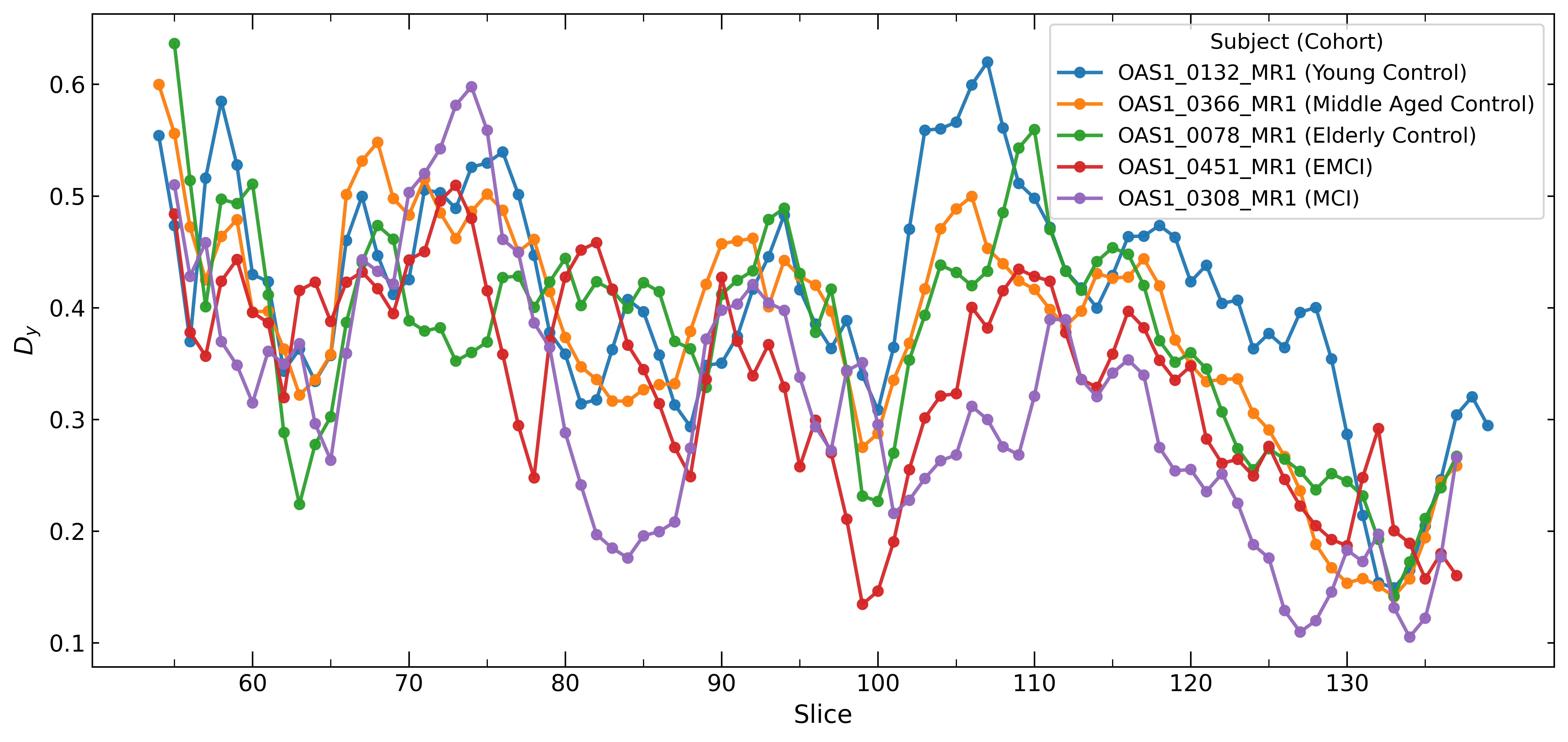}
    \caption{$\Delta\alpha_y$ slice profiles for the OASIS-1 for representative individuals from each cohort. Sample fluctuation functions for the same individuals are presented in Fig.~\ref{fig:fluctuatio_funtion_sample} in the main text.}
    \label{fig:example_subject_profiles}
\end{figure}

To illustrate the features of fluctuation functions and multifractal spectra that undergo changes in aging and disease (shown in ~\ref{sec:cascades} of the main text), we selected  a representative individual from each cohort. These individuals were hand-picked so that their $\Delta\alpha_y$ slice profile follows the cohort median. 
The slice profiles of the selected subjects are presented  in figure~\ref{fig:example_subject_profiles}.

\subsection{LDA interpretation}

The PCA-LDA pipeline was performed with 50 most important principal components (reducing dimensionality from 777 features) and then 2 linear discriminants.
In the LD space, the samples follow roughly a U-shape starting with the young control group, followed by the middle-aged control, elderly control, EMCI and finally MCI groups.
It then stands to reason, to analyze feature differences only between the consecutive groups (and not their all pairs).

Each row in Fig.~\ref{supp_fig:raw_prototypes} below shows a difference between means of feature intensity in consecutive groups.
All available features are presented: $H$, $A$ and $\Delta\alpha$ metrics on each slice of all three axes.
This plot reveals among others that multifractal spectrum width in slices around 65 and 115 on x-axis (and similarly Hurst exponent around slice 100 on z-axis) very strongly differs between the middle aged group and the elderly and then, progressively, a similar difference can be found between the elderly and EMCI and between EMCI and MCI subjects.
In this respect, the neurodegenerative changes could be interpreted as an extension of the aging process.
The strongest differences happen between the middle-aged and the elderly, while changes between the other consecutive pairs are lower.

\begin{figure}[ht]
    \centering
    \includegraphics[scale=0.4]{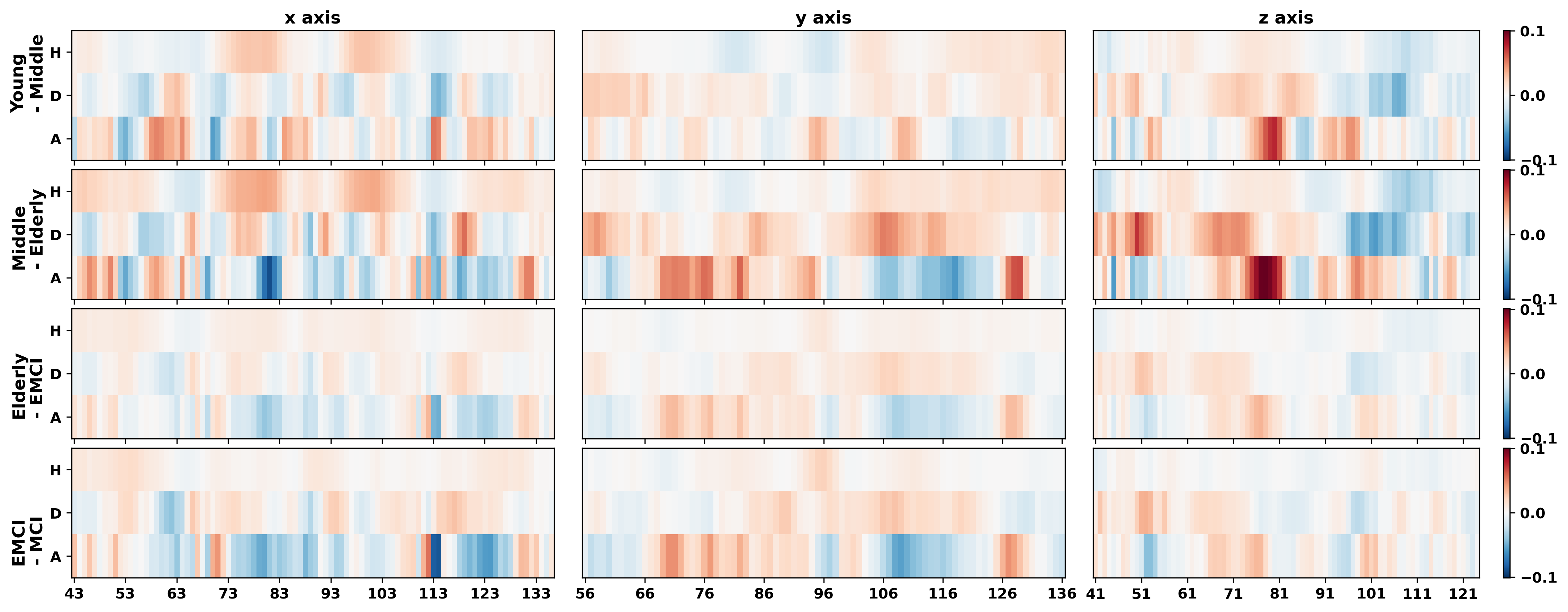}
    \caption{Features reconstructed from differences between group centroids in LD space.}
    \label{supp_fig:LDA_prototypes}
\end{figure}

Now, the linear discriminant space serves as a denoised feature space maximizing discrimination between the groups. The $H$, $A$ and $\Delta\alpha$ features can be reconstructed by taking pseudo-inverse of the LDA projection matrix and reversing the centering in PCA space.
To find the group differences we thus determine group centroids in LD space, compute their differences and perform the reconstruction to the original feature space, which is presented in Fig.~\ref{supp_fig:LDA_prototypes}.
The picture changes with respect to the raw feature values. For instance, (i) features with high intensity changes appear in all group differences (not just middle-aged vs the elderly); (ii) Hurst exponent is mostly useful in discrimination between the young and the middle-aged and the middle-aged and the elderly, but not for cognitive impairment progression; (iii) the features most strongly discriminating between the middle-aged and the elderly, but also further the elderly from EMCI and EMCI from MCI, are asymmetry features (e.g., around slices 70-80 and 125-130 on y axis and slices around 70-80 on z axis); (iv) the z-axis asymmetry difference seems to progresively migrate from slices around 80 to slices around 70, as if the location of the change was age- and disease-related.
The general observation is that the multifractal features strengthen the discrimination even though the differences in raw values are smaller in magnitude.

\begin{figure}[ht]
    \centering
    \includegraphics[scale=0.4]{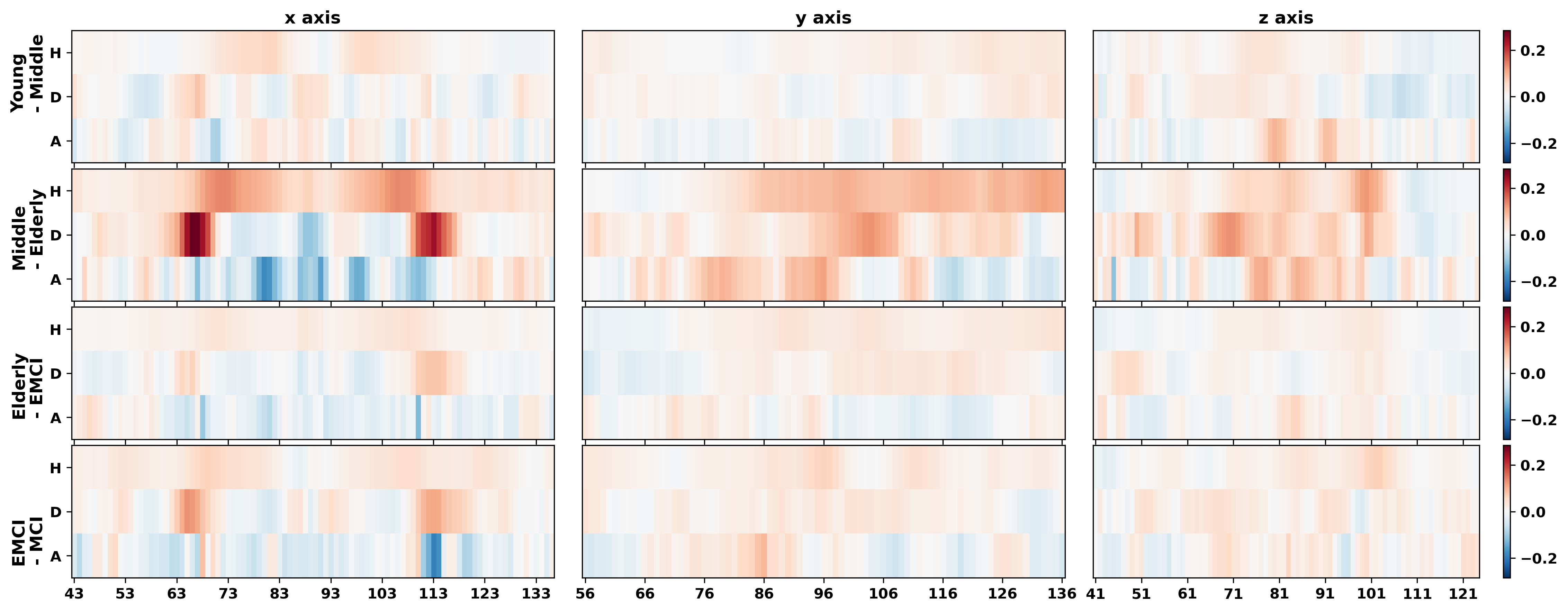}
    \caption{Differences between features of group means.}
    \label{supp_fig:raw_prototypes}
\end{figure}

\begin{thebibliography}{10}

\bibitem{alzheimer2019world}
Alzheimer's~Disease International.
\newblock {\em World {A}lzheimer {R}eport 2019: {A}ttitudes to dementia}.
\newblock London, England: Alzheimer's Disease International, 2019.

\bibitem{nichols2022estimation}
Emma Nichols, Jaimie~D Steinmetz, Stein~Emil Vollset, Kai Fukutaki, Julian Chalek, Foad Abd-Allah, Amir Abdoli, Ahmed Abualhasan, Eman Abu-Gharbieh, Tayyaba~Tayyaba Akram, et~al.
\newblock Estimation of the global prevalence of dementia in 2019 and forecasted prevalence in 2050: an analysis for the {G}lobal {B}urden of {D}isease {S}tudy 2019.
\newblock {\em The Lancet Public Health}, 7(2):e105--e125, 2022.

\bibitem{alzheimer2021world}
Serge Gauthier, Pedro Rosa-Neto, Jos{\'e}~A Morais, and Claire Webster.
\newblock {\em World {A}lzheimer {R}eport 2021: {J}ourney through the diagnosis of dementia}.
\newblock London, England: Alzheimer’s Disease International, 2021.

\bibitem{jack2018nia}
Clifford~R Jack~Jr, David~A Bennett, Kaj Blennow, Maria~C Carrillo, Billy Dunn, Samantha~Budd Haeberlein, David~M Holtzman, William Jagust, Frank Jessen, Jason Karlawish, et~al.
\newblock {NIA-AA} research framework: toward a biological definition of {A}lzheimer's disease.
\newblock {\em Alzheimer's \& dementia}, 14(4):535--562, 2018.

\bibitem{jack2024revised}
Clifford~R Jack~Jr, J~Scott Andrews, Thomas~G Beach, Teresa Buracchio, Billy Dunn, Ana Graf, Oskar Hansson, Carole Ho, William Jagust, Eric McDade, et~al.
\newblock Revised criteria for diagnosis and staging of {A}lzheimer's disease: {A}lzheimer's {A}ssociation {W}orkgroup.
\newblock {\em Alzheimer's \& Dementia}, 20(8):5143--5169, 2024.

\bibitem{sperling2011toward}
Reisa~A Sperling, Paul~S Aisen, Laurel~A Beckett, David~A Bennett, Suzanne Craft, Anne~M Fagan, Takeshi Iwatsubo, Clifford~R Jack~Jr, Jeffrey Kaye, Thomas~J Montine, et~al.
\newblock Toward defining the preclinical stages of {A}lzheimer’s disease: Recommendations from the {N}ational {I}nstitute on {A}ging-{A}lzheimer's {A}ssociation workgroups on diagnostic guidelines for {A}lzheimer's disease.
\newblock {\em Alzheimer's \& dementia}, 7(3):280--292, 2011.

\bibitem{aschenbrenner2022comparison}
Andrew~J Aschenbrenner, Yan Li, Rachel~L Henson, Katherine Volluz, Jason Hassenstab, Philip Verghese, Tim West, Matthew~R Meyer, Kristopher~M Kirmess, Anne~M Fagan, et~al.
\newblock Comparison of plasma and {CSF} biomarkers in predicting cognitive decline.
\newblock {\em Annals of Clinical and Translational Neurology}, 9(11):1739--1751, 2022.

\bibitem{therriault2023equivalence}
Joseph Therriault, Stijn Servaes, C{\'e}cile Tissot, Nesrine Rahmouni, Nicholas~J Ashton, Andr{\'e}a~Lessa Benedet, Thomas~K Karikari, Arthur~C Macedo, Firoza~Z Lussier, Jenna Stevenson, et~al.
\newblock Equivalence of plasma p-tau217 with cerebrospinal fluid in the diagnosis of {A}lzheimer's disease.
\newblock {\em Alzheimer's \& Dementia}, 19(11):4967--4977, 2023.

\bibitem{noda2024cost}
Kenta Noda, Yeongjoo Lim, Rei Goto, Shintaro Sengoku, and Kota Kodama.
\newblock Cost-effectiveness comparison between blood biomarkers and conventional tests in {A}lzheimer's disease diagnosis.
\newblock {\em Drug Discovery Today}, 29(3):103911, 2024.

\bibitem{scholl2025cutting}
M~Sch{\"o}ll, A~Vrillon, T~Ikeuchi, FC~Quevenco, L~Iaccarino, SZ~Vasileva-Metodiev, SC~Burnham, J~Hendrix, S~Epelbaum, H~Zetterberg, et~al.
\newblock Cutting through the noise: a narrative review of {A}lzheimer's disease plasma biomarkers for routine clinical use.
\newblock {\em The Journal of Prevention of Alzheimer's Disease}, page 100056, 2025.

\bibitem{kubota2025plasma}
Masahito Kubota, Shogyoku Bun, Keisuke Takahata, Shin Kurose, Yuki Momota, Yu~Iwabuchi, Toshiki Tezuka, Hajime Tabuchi, Morinobu Seki, Yasuharu Yamamoto, et~al.
\newblock Plasma biomarkers for early detection of {A}lzheimer’s disease: a cross-sectional study in a {J}apanese cohort.
\newblock {\em Alzheimer's Research \& Therapy}, 17(1):131, 2025.

\bibitem{palmqvist2025plasma}
Sebastian Palmqvist, No{\"e}lle Warmenhoven, Federica Anastasi, Andrea Pilotto, Shorena Janelidze, Pontus Tideman, Erik Stomrud, Niklas Mattsson-Carlgren, Ruben Smith, Rik Ossenkoppele, et~al.
\newblock Plasma phospho-tau217 for {A}lzheimer’s disease diagnosis in primary and secondary care using a fully automated platform.
\newblock {\em Nature Medicine}, pages 1--8, 2025.

\bibitem{dang2023neuroimaging}
Chun Dang, Yanchao Wang, Qian Li, and Yaoheng Lu.
\newblock Neuroimaging modalities in the detection of {A}lzheimer's disease-associated biomarkers.
\newblock {\em Psychoradiology}, 3:kkad009, 2023.

\bibitem{greicius2004default}
Michael~D Greicius, Gaurav Srivastava, Allan~L Reiss, and Vinod Menon.
\newblock Default-mode network activity distinguishes {A}lzheimer's disease from healthy aging: evidence from functional {MRI}.
\newblock {\em Proceedings of the National Academy of Sciences}, 101(13):4637--4642, 2004.

\bibitem{wu2011altered}
Xia Wu, Rui Li, Adam~S Fleisher, Eric~M Reiman, Xiaoting Guan, Yumei Zhang, Kewei Chen, and Li~Yao.
\newblock Altered default mode network connectivity in {A}lzheimer's disease—a resting functional {MRI} and {B}ayesian network study.
\newblock {\em Human brain mapping}, 32(11):1868--1881, 2011.

\bibitem{schuff2009mri}
Norbert Schuff, N~Woerner, L~Boreta, T~Kornfield, LM~Shaw, JQ~Trojanowski, PM~Thompson, CR~Jack~Jr, MW~Weiner, and Alzheimer's; Disease~Neuroimaging Initiative.
\newblock {MRI} of hippocampal volume loss in early {A}lzheimer's disease in relation to {ApoE} genotype and biomarkers.
\newblock {\em Brain}, 132(4):1067--1077, 2009.

\bibitem{struyfs2020automated}
Hanne Struyfs, Diana~Maria Sima, Melissa Wittens, Annemie Ribbens, Nuno~Pedrosa de~Barros, Thanh V{\^a}n~Phan, Maria Ines~Ferraz Meyer, Lene Claes, Ellis Niemantsverdriet, Sebastiaan Engelborghs, et~al.
\newblock Automated {MRI} volumetry as a diagnostic tool for {A}lzheimer's disease: Validation of icobrain dm.
\newblock {\em NeuroImage: Clinical}, 26:102243, 2020.

\bibitem{Bachli2020EvaluatingApproach}
M.B. Bachli, L.~Sede{\~{n}}o, J.K. Ochab, O.~Piguet, F.~Kumfor, P.~Reyes, T.~Torralva, M.~Roca, J.F. Cardona, C.G. Campo, E.~Herrera, A.~Slachevsky, D.~Matallana, F.~Manes, A.M. Garc{\'{i}}a, A.~Ib{\'{a}}{\~{n}}ez, and D.R. Chialvo.
\newblock {Evaluating the reliability of neurocognitive biomarkers of neurodegenerative diseases across countries: A machine learning approach}.
\newblock {\em NeuroImage}, 208, 2020.

\bibitem{doering2024mri}
Elena Doering, Georgios Antonopoulos, Merle Hoenig, Thilo van Eimeren, Marcel Daamen, Henning Boecker, Frank Jessen, Emrah D{\"u}zel, Simon Eickhoff, Kaustubh Patil, et~al.
\newblock {MRI} or {18F-FDG} {PET} for brain age gap estimation: links to cognition, pathology, and {Alzheimer} disease progression.
\newblock {\em Journal of nuclear medicine}, 65(1):147--155, 2024.

\bibitem{karim2024identifying}
SM~Shayez Karim, Md~Shah Fahad, and RS~Rathore.
\newblock Identifying discriminative features of brain network for prediction of {A}lzheimer’s disease using graph theory and machine learning.
\newblock {\em Frontiers in neuroinformatics}, 18:1384720, 2024.

\bibitem{aghdam2025machine}
Maryam~Akhavan Aghdam, Serdar Bozdag, and Fahad Saeed.
\newblock Machine-learning models for {A}lzheimer’s disease diagnosis using neuroimaging data: survey, reproducibility, and generalizability evaluation.
\newblock {\em Brain Informatics}, 12(1):1--27, 2025.

\bibitem{ziukelis2022fractal}
Elina~T Ziukelis, Elijah Mak, Maria-Eleni Dounavi, Li~Su, and John T~O'Brien.
\newblock Fractal dimension of the brain in neurodegenerative disease and dementia: a systematic review.
\newblock {\em Ageing research reviews}, 79:101651, 2022.

\bibitem{eke2012pitfalls}
Andras Eke, Peter Herman, Basavaraju~G Sanganahalli, Fahmeed Hyder, Peter Mukli, and Zoltan Nagy.
\newblock Pitfalls in fractal time series analysis: {fMRI} {BOLD} as an exemplary case.
\newblock {\em Frontiers in physiology}, 3:417, 2012.

\bibitem{rohini2020differentiation}
P~Rohini, S~Sundar, and S~Ramakrishnan.
\newblock Differentiation of early mild cognitive impairment in brainstem mr images using multifractal detrended moving average singularity spectral features.
\newblock {\em Biomedical Signal Processing and Control}, 57:101780, 2020.

\bibitem{long2018brainnetome}
Zhuqing Long, Bin Jing, Ru~Guo, Bo~Li, Feiyi Cui, Tingting Wang, and Hongwen Chen.
\newblock A brainnetome atlas based mild cognitive impairment identification using {H}urst exponent.
\newblock {\em Frontiers in aging neuroscience}, 10:103, 2018.

\bibitem{long2023identifying}
Zhuqing Long, Jie Li, Jianghua Fan, Bo~Li, Yukeng Du, Shuang Qiu, Jichang Miao, Jian Chen, Juanwu Yin, and Bin Jing.
\newblock Identifying {A}lzheimer’s disease and mild cognitive impairment with atlas-based multi-modal metrics.
\newblock {\em Frontiers in Aging Neuroscience}, 15:1212275, 2023.

\bibitem{reishofer2018age}
Gernot Reishofer, Fritz Studencnik, Karl Koschutnig, Hannes Deutschmann, Helmut Ahammer, and Guilherme Wood.
\newblock Age is reflected in the fractal dimensionality of mri diffusion based tractography.
\newblock {\em Scientific reports}, 8(1):5431, 2018.

\bibitem{madan2016cortical}
Christopher~R Madan and Elizabeth~A Kensinger.
\newblock Cortical complexity as a measure of age-related brain atrophy.
\newblock {\em NeuroImage}, 134:617--629, 2016.

\bibitem{madan2018predicting}
Christopher~R Madan and Elizabeth~A Kensinger.
\newblock Predicting age from cortical structure across the lifespan.
\newblock {\em European Journal of Neuroscience}, 47(5):399--416, 2018.

\bibitem{farahibozorg2015age}
S~Farahibozorg, SM~Hashemi-Golpayegani, and J~Ashburner.
\newblock Age-and sex-related variations in the brain white matter fractal dimension throughout adulthood: an mri study.
\newblock {\em Clinical neuroradiology}, 25(1):19--32, 2015.

\bibitem{krohn2026fractal}
Stephan Krohn, Amy Romanello, Nina Von~Schwanenflug, Jerod~M Rasmussen, Claudia Buss, Sofie~L Valk, Christopher~R Madan, and Carsten Finke.
\newblock Fractal analysis of brain shape formation predicts age and genetic similarity in human newborns.
\newblock {\em Nature Neuroscience}, 29(1):171--185, 2026.

\bibitem{dong2018hurst}
Jianxin Dong, Bin Jing, Xiangyu Ma, Han Liu, Xiao Mo, and Haiyun Li.
\newblock Hurst exponent analysis of resting-state fmri signal complexity across the adult lifespan.
\newblock {\em Frontiers in neuroscience}, 12:34, 2018.

\bibitem{stanyard2024aperiodic}
RA~Stanyard, David Mason, Claire Ellis, Hannah Dickson, Roxanna Short, Dafnis Batalle, and Tomoki Arichi.
\newblock Aperiodic and hurst eeg exponents across early human brain development: A systematic review.
\newblock {\em Developmental Cognitive Neuroscience}, 68:101402, 2024.

\bibitem{berthouze2010human}
Luc Berthouze, Leon~M James, and Simon~F Farmer.
\newblock Human eeg shows long-range temporal correlations of oscillation amplitude in theta, alpha and beta bands across a wide age range.
\newblock {\em Clinical Neurophysiology}, 121(8):1187--1197, 2010.

\bibitem{davoudi2025electroencephalography}
Saeideh Davoudi, Gabriela~Lopez Arango, Florence Deguire, Inga~Sophie Knoth, Fanny Thebault-Dagher, Rebecca Reh, Laurel Trainor, Janet Werker, and Sarah Lipp{\'e}.
\newblock Electroencephalography estimates brain age in infants with high precision: Leveraging advanced machine learning in healthcare.
\newblock {\em NeuroImage}, 312:121200, 2025.

\bibitem{cauzzo2026detrended}
Simone Cauzzo, Sadaf Moaveninejad, Angelo Antonini, Maurizio Corbetta, and Camillo Porcaro.
\newblock Detrended fluctuation analysis complements spectral features in characterizing functional brain aging.
\newblock {\em Fractal and Fractional}, 10(4):224, 2026.

\bibitem{grela2025using}
Jacek Grela, Zbigniew Drogosz, Jakub Janarek, Jeremi~K. Ochab, Ignacio Cifre, Ewa Gudowska-Nowak, Maciej~A. Nowak, Pawe{\l} O{\'s}wi{\k{e}}cimka, and Dante~R. Chialvo.
\newblock Using space-filling curves and fractals to reveal spatial and temporal patterns in neuroimaging data.
\newblock {\em Journal of Neural Engineering}, 22(1):016016, 2025.

\bibitem{bader2012spacefilling}
Michael Bader.
\newblock {\em Space-filling curves: an introduction with applications in scientific computing}.
\newblock Springer, 2012.

\bibitem{kantelhardt2002multifractal}
Jan~W. Kantelhardt, Stephan~A. Zschiegner, Eva Koscielny-Bunde, Shlomo Havlin, Armin Bunde, and H.~Eugene Stanley.
\newblock Multifractal detrended fluctuation analysis of nonstationary time series.
\newblock {\em Physica A: Statistical Mechanics and its Applications}, 316(1--4):87--114, 2002.

\bibitem{OASIS1}
Daniel~S Marcus, Tracy~H Wang, Jamie Parker, John~G Csernansky, John~C Morris, and Randy~L Buckner.
\newblock {Open Access Series of Imaging Studies (OASIS)}: cross-sectional {MRI} data in young, middle aged, nondemented, and demented older adults.
\newblock {\em J Cogn Neurosci}, 19(9):1498--1507, September 2007.

\bibitem{wustl_isp_2000}
{Washington University Memory and Aging Project}.
\newblock {\em Initial Subject Protocol (ISP)}.
\newblock Washington University in St.\ Louis, 2000.
\newblock © December 2000; Major Revision 01/98. Includes the Hollingshead Index of Social Position scoring worksheet (see early pages of the PDF).

\bibitem{vandermaaten2008tsne}
Laurens van~der Maaten and Geoffrey Hinton.
\newblock Visualizing data using t-sne.
\newblock {\em Journal of Machine Learning Research}, 9:2579--2605, 2008.

\bibitem{theodoridis2009Pattern}
Sergios Theodoridis and Konstantinos Koutroumbas.
\newblock {\em Pattern Recognition}.
\newblock Academic Press, Burlington, MA London, 4th ed edition, 2009.

\bibitem{Cheng2005}
Qiuming Cheng.
\newblock Multifractal distribution of eigenvalues and eigenvectors from 2d multiplicative cascade multifractal fields.
\newblock {\em Mathematical Geology}, 37(8):915--927, 2005.

\bibitem{Xu2017}
Hai-Chuan Xu, Gao-Feng Gu, and Wei-Xing Zhou.
\newblock A direct determination approach for the multifractal detrending moving average analysis.
\newblock {\em Physical Review E}, 96(5):052201, 2017.

\bibitem{ADNI}
R~C Petersen, P~S Aisen, L~A Beckett, M~C Donohue, A~C Gamst, D~J Harvey, C~R Jack, Jr, W~J Jagust, L~M Shaw, A~W Toga, J~Q Trojanowski, and M~W Weiner.
\newblock {Alzheimer's Disease Neuroimaging Initiative (ADNI)}: clinical characterization.
\newblock {\em Neurology}, 74(3):201--209, December 2009.

\bibitem{Poldrack2015}
Timothy O. Laumann, Evan M. Gordon, Babatunde Adeyemo, Abraham Z. Snyder, Sung Jun Joo, Mei-Yen Chen, Adrian W. Gilmore, Kathleen B. McDermott, Steven M. Nelson, Nico U.F. Dosenbach, Bradley L. Schlaggar, Jeanette A. Mumford, Russell A. Poldrack, and Steven E. Petersen.
\newblock Functional system and areal organization of a highly sampled individual human brain.
\newblock {\em Neuron}, 87(3):657--670, Aug 2015.

\end{thebibliography}
\end{document}